\documentclass[10pt,a4paper,twoside]{article}

\usepackage{mathtools,amssymb,amsthm}						%Standard math stuff

\usepackage{hyperref}
	\hypersetup{colorlinks,linkcolor={red!50!black},citecolor={blue!50!black},urlcolor={blue!80!black}}

\usepackage[backend=biber,style=authoryear-icomp,natbib=true,
    sortcites=true,ibidtracker=strict,sorting=nyvt,giveninits=true,
    uniquename=init,maxbibnames=99]{biblatex}
\renewbibmacro{in:}{} %Removes "In:" before journal name
\newbibmacro{string+doi}[1]{%
    \iffieldundef{doi}{#1}{\href{http://dx.doi.org/\thefield{doi}}{#1}}}
\DeclareFieldFormat{title}{\usebibmacro{string+doi}{\mkbibemph{#1}}}
\DeclareFieldFormat[article]{title}{\usebibmacro{string+doi}{\mkbibquote{#1}}}

\usepackage[top=3cm,bottom=2cm,inner=2cm,outer=2cm,a4paper]{geometry}

\usepackage{tikz}

\usepackage[font=small,labelfont=bf,centerlast]{caption}

\usepackage{calc}
\usepackage{graphicx} %or add to packages_and_options
\usepackage{xcolor}
\newcommand{\ncom}{\newcommand}

\ncom{\h}{\mathcal{H}}
\ncom{\ket}[1]{\left|#1\right\rangle}
\ncom{\bra}[1]{\left\langle#1\right|}
\ncom{\braket}[2]{\left\langle#1\middle|#2\right\rangle}
\ncom{\ketbra}[2]{\left|#1\middle\rangle\middle\langle#2\right|}
\ncom{\mean}[1]{\left\langle#1\right\rangle}
\ncom{\expv}[3]{\left\langle#1\middle|#2\middle|#3\right\rangle}

\ncom{\set}[2]{\left\{#1\:\middle|\:#2\right\}}				%typesetting set
\ncom{\dee}{\,\mathrm{d}}							%straight d for infinitesimals
\ncom{\ldef}{\coloneqq}								%:= with correct positioning
\ncom{\rdef}{\eqqcolon}								%=: with correct positioning

\DeclareMathOperator{\supp}{\mathrm{supp}}				%Support

\title{Extending Bell's Theorem: Nonlocality via Measurement Dependence}
\author{G. Bacciagaluppi,\thanks{Descartes Centre for the History and Philosophy of the Sciences and the Humanities and Freudenthal Institute, Utrecht University; IHPST and SPHERE, Paris; Foundational Questions Institute, \url{https://fqxi.org/} (email: g.bacciagaluppi@uu.nl).}\,  R. Hermens\thanks{Freudenthal Institute, Utrecht University (email: r.hermens@uu.nl).}\, and G. Leegwater\thanks{Erasmus School of Philosophy, Erasmus University Rotterdam (email:  leegwater@esphil.eur.nl).}}
\date{4 July 2026}

\begin{document}

\maketitle

\begin{abstract} 
Besides well-known conditions of locality or factorisability, deriving the Bell inequalities requires assuming that the distribution of hidden variables and Alice's and Bob's measurement settings be independent of each other. We show that (analogously to violations of locality due to action at a distance) certain violations of this Measurement Independence assumption can be associated with a notion of signalling in principle, thus making them also testable in principle, and spell out the appropriate conditions. Accordingly, we show that by imposing no-signalling one can prove a version of Bell's theorem that does not require the assumption of Measurement Independence. We discuss the `Schulman model' as an example, as well as lessons for `experimental metaphysics'.
\end{abstract}

\tableofcontents

%\input{command_examples}

%%%%%%%%%%%%%%%%%%%%%%%
%                                                               %
%%%%%%%%%%%%%%%%%%%%%%%
\section{Introduction}\label{intro}
Many standard analyses of Bell's theorem emphasise the two assumptions that Abner Shimony called Outcome Independence (OI) and Parameter Independence (PI), or very similar ones. They conclude from the (loophole-free) experimental violation of the Bell inequalities  that at least one of these assumptions must be violated, resulting in some form of nonlocality (which may be incompatible with or merely in tension with relativity).\footnote{There are several derivations of the Bell inequalities, so their experimental violation refutes simultaneously several sets of assumptions. The analysis by Shimony (1986) is based on derivations that assume Bell's factorisability condition (\ref{2b}) below. Other analyses are phrased in terms of somewhat stronger versions of `local realism', where the `locality' assumption is essentially PI and the `realism' assumption is usually either a form of separability or the existence of deterministic or non-contextual hidden variables. Derivations based on factorisabilty have the advantage of not requiring any `realism' assumptions beyond the existence of \emph{some} initial state $\lambda$ of the total system.} There are some approaches to quantum mechanics, however, that aim to explain the violation of the Bell inequalities in \emph{local} terms, by rejecting the third assumption needed in the derivations,  known as Measurement Independence (MI), namely that Alice and Bob's measurement settings and the distribution of hidden variables at the source are independent of each other.\footnote{This assumption is often called also Settings Independence or Settings--Source Independence.} Recent analyses have further shown that if one quantifies the dependence between the hidden variables distribution and the measurement settings, then in the standard case of binary observables only small violations of MI are needed to reproduce the experimental violations of the Bell inequalities, so that, conversely, it is possible to derive the inequalities, even allowing for violations of MI, by limiting how much the hidden variables distributions depend on the settings.\footnote{See Barrett and Gisin (2011), Hall (2011), Friedman \emph{et al.}\ (2019), Vieira \emph{et al.}\ (2025). Note that for the general case with observables with an arbitrary number of possible outcomes, this is no longer true; see again Vieira \emph{et al.}\ (2025).}

This paper proposes a different perspective on MI and its possible violations. We will show that one can associate a notion of `signalling in principle' also with some (not all) violations of MI, and we formulate conditions under which such signalling in principle would manifest itself operationally, making these violations testable. Some of these conditions are implied by the violation of the Bell inequalities, and we extend Bell's theorem accordingly to models that may violate MI. While we agree with the claims that such signalling in principle is compatible with relativity (because a signal would proceed along a zig-zag path backwards and forwards in spacetime), it constitutes a new form of nonlocality, and we provide an analysis of Bell's theorem in terms of the different and in principle operationally distinguishable forms of nonlocality associated with violations of OI, PI \emph{and} MI. 

The paper is structured as follows. We begin in Section~\ref{neglected} by reviewing in more detail the assumption of MI in Bell's theorem. In Section~\ref{signalling} we make precise what we mean by `signalling in principle', and in Section~\ref{conditions} we construct an explicit signalling protocol covering also violations of MI. Then in Section~\ref{strengthening} we turn to Bell's theorem and extend it by relaxing the assumption of MI as just indicated. Section~\ref{limitations} presents a detailed example (the `Schulman model'), and Section~\ref{metaphysics} analyses the resulting nonlocality. Technical proofs are postponed to three Appendices.

%%%%%%%%%%%%%%%%%%%%%%%
%                                                               %
%%%%%%%%%%%%%%%%%%%%%%%
\section{Measurement Independence}\label{neglected}
Take a standard Bell experiment setup: Charlie prepares an ensemble of spin-$\frac{1}{2}$ pairs in a quantum state $\rho$, and Alice and Bob measure spin variables $A$ or $A'$ and $B$ or $B'$ (with values $\pm1$). Their measurement contexts will be denoted as $I$ or $I'$ and $J$ or $J'$.\footnote{Such contexts might include in particular the polarity and gradient of the magnetic fields of their Stern--Gerlach apparatuses, on which trajectories in the Bohm theory depend (see Section~\ref{signalling}).} Assume further that each such setup defines a statistical distribution $\sigma_\rho^{IJ}(\lambda)$ or $ \sigma_\rho^{I'J}(\lambda)$ or $\sigma_\rho^{IJ'}(\lambda)$ or $\sigma_\rho^{I'J'}(\lambda)$ for some appropriate variables $\lambda$ denoting the state of the system. The individual $\lambda$ in turn define probabilities for outcomes of measurements of pairs of quantities, say $p_\lambda^{IJ}(A=\pm 1,B=\pm 1)$. We write 
  \begin{equation}
      \mean{AB}_\lambda^{IJ}\ldef p_\lambda^{IJ}(A=B)-p_\lambda^{IJ}(A\neq B)
    \label{0}
  \end{equation}
and
  \begin{equation}
      \mean{AB}_\rho^{IJ}\ldef 
      \int_{\supp(\sigma_\rho^{IJ})}\sigma_\rho^{IJ}(\lambda)d\lambda\mean{AB}_\lambda^{IJ} \ .
    \label{0b}
  \end{equation}
We shall follow a fairly common usage and call this a `hidden variables' model, but $\lambda$ will generally include both the quantum state of the system and any additional variables that may be needed to specify its state.\footnote{In this sense, quantum mechanics itself is a trivial `hidden variables' theory.}

The Bell inequalities can now be derived as follows.\footnote{We shall always use the Bell inequalities in CHSH form.} First, for all combinations of primed and unprimed quantities and for almost all hidden variables in the support of the distributions $\sigma_\rho^{IJ}(\lambda)$ etc.\  (i.e.\ up to sets of measure zero) assume both \emph{Outcome Independence} (OI),
  \begin{equation}
    \mean{AB}^{IJ}_\lambda = \mean{A}^{IJ}_\lambda \mean{B}^{IJ}_\lambda 
    \label{1}
  \end{equation}
(where the expressions of the form $\mean{A}^{IJ}_\lambda$ denote simply the marginal expectation values), and \emph{Parameter Independence} (PI),
  \begin{equation}
    \mean{A}^{IJ}_\lambda = \mean{A}^{I}_\lambda\ ,\qquad 
    \mean{B}^{IJ}_\lambda = \mean{B}^{J}_\lambda 
    \label{2}
  \end{equation} 
(meaning that the left-hand sides in fact do not depend on $J$ or $I$, respectively). The conjunction of OI and PI gives Bell's \emph{factorisability condition}
  \begin{equation}
     \mean{AB}^{IJ}_\lambda = \mean{A}^{I}_\lambda\mean{B}^{J}_\lambda \ .
     \label{2b}
  \end{equation}
Then assume \emph{Measurement Independence} (MI), i.e.\ the distribution of the hidden variables should be independent of $I,I',J,J'$:
  \begin{equation}
     \sigma_\rho(\lambda)\ldef\sigma_\rho^{IJ}(\lambda)=\sigma_\rho^{I'J}(\lambda)=\sigma_\rho^{IJ'}(\lambda)=\sigma_\rho^{I'J'}(\lambda)\ ,
     \label{5}
  \end{equation}
where the right-hand side means that (up to sets of measure zero) the four distributions have the same support and are equal. Under this assumption we can consider the combination
  \begin{equation}
    \mean{AB}^{IJ}_\lambda  + \mean{A'B}^{I'J}_\lambda  + \mean{AB'}^{IJ'}_\lambda  - \mean{A'B'}^{I'J'}_\lambda  
    \label{3}
  \end{equation}
(and similarly with the minus sign positioned differently), and average it over $\sigma_\rho(\lambda)$. But by (\ref{2b}), expression (\ref{3}) is equal to  
 \begin{equation}
    \mean{A}^{I}_\lambda\mean{B}^{J}_\lambda + 
    \mean{A'}^{I'}_\lambda\mean{B}^{J}_\lambda + 
    \mean{A}^{I}_\lambda\mean{B'}^{J'}_\lambda - 
    \mean{A'}^{I'}_\lambda\mean{B'}^{J'}_\lambda \ ,   
    \label{4}
  \end{equation}
which always lies in $[-2,+2]$ for $\pm 1$-valued quantities. Thus, averaging it over $\sigma_\rho(\lambda)$ also yields a result in $[-2,+2]$, and we have derived the \emph{Bell inequalities}:
 \begin{equation}
    -2\leq\mean{AB}_\rho^{IJ} + 
    \mean{A'B}_\rho^{I'J} + 
    \mean{AB'}_\rho^{IJ'} - 
    \mean{A'B'}_\rho^{I'J'}\leq 2    
    \label{6}
  \end{equation}
(and similarly with the minus sign positioned differently). Without assuming MI, even if OI and PI are satisfied, each of the terms in (\ref{6}) is an average over a separate distribution, and we can only say (trivially) that the whole expression lies in $[-4,+4]$, so that the Bell inequalities can be violated.\footnote{For further discussion see Hermens (2019).}

The interest in MI lies in the potential for \emph{local explanations} of such violations: one imagines that neither OI or PI are violated (whatever form of nonlocality they correspond to exactly) and that the hidden variables $\lambda$ affect Alice and Bob's outcomes purely locally. But if their distribution is appropriately different in runs of the experiment in which Alice and Bob measure different pairs of spin quantities, the Bell inequalities will in fact be violated.

The obvious question is how such appropriate dependence between the hidden variables and the measurement settings should come about. Bell among others calls violations of MI `conspiratorial' (Bell 1981). If one imagines that the dependence between the hidden variables and the measurement settings is due to some contingent cause (in their common past), it would indeed be a conspiracy that such dependences are created independently of the type of process that fixes the settings (whether human choices, optical switches, Swiss lottery machines or distant quasars). If there are no systematic mechanisms that can exploit such local common causes, and one thus equates common-cause explanations with conspiracies,\footnote{Some authors use the term `conspiratorial' more broadly to include theories where the initial conditions are fine-tuned in some appropriate sense. For instance, Sen and Valentini (2020a,b) argue that all deterministic models violating MI are conspiratorial in this broader sense. (We do not consider such fine-tuning to be a problem in general, as we explain in Section~\ref{metaphysics} below.)} it comes as no surprise that these explanations can be largely ruled out experimentally, much like other experimental loopholes in the Bell experiments such as the `detection loophole' and the `locality loophole'.\footnote{The detection (or `efficiency' of `fair-sampling') loophole is based on the idea that the hidden variables might instruct certain particles to avoid detection, so that the samples of particles that are actually detected are unrepresentative and may contain spurious correlations. This loophole requires that a large enough fraction of particles go undetected, and has been ruled out by increasing the observed efficiency of the detectors. The locality loophole invokes the possibility of subluminal mechanisms transferring information between various parts of the experiment (e.g.\ from Alice's to Bob's lab), and has been ruled out by ensuring that they are spacelike separated. Finally, common-cause explanations cannot be ruled out completely, but have been made hugely implausible by experiments such as the `Big Bell' (where the settings were determined by thousands of citizen scientists) and the `Cosmic Bell' (where the settings were determined by light from distant quasars). See Kaiser (2022) for an excellent overview.} On the other hand, just as closing the detection loophole does not rule out genuine violations of OI (e.g.\ collapse) or closing the locality loophole does not rule out genuine violations of PI (e.g.\ Bohm), closing this `conspiracy loophole' (or `common-cause loophole') does not rule out \emph{genuine violations of MI} such as may be due, say, to direct retrocausal influences of the settings on the source or to global constraints that enforce dependence.\footnote{The idea of retrocausality has been vigorously championed in particular by Price (1996), and is often investigated in the context of theories specifying boundary conditions not only in the past but also in the future, as pursued by Wharton and co-workers (see for instance Wharton (2014), Almada \emph{et al.}\ (2016), Wharton \emph{et al.}\ (2024)). Dependence between settings and hidden variables may also be enforced by global constraints that get propagated by the dynamics, as in the cellular automaton theory by 't~Hooft (2016), or by the structure of the state space, as in the invariant set theory by Palmer (2020). Note that if one takes causation as an emergent notion (which we are sympathetic to), the distinction between retrocausal and other ways of genuinely violating MI may be blurred. In the deterministic case, one often describes theories that violate MI as `superdeterministic' (because the settings are not taken to be free variables), although terminology varies (e.g.\ Vervoort (2013) uses `superdeterminism' to refer to common-cause models, while Hossenfelder and Palmer (2020) group together under this label the approaches by Wharton and co-authors, by 't~Hooft and by Palmer. We emphasise that determinism is not required for violations of MI, especially given that, as mentioned, only a small amount of dependence is needed to violate the Bell inequalities. There are other possible candidates for theories violating MI, often formulated in an indeterministic setting, namely ones that postulate a highly correlated background field that interacts with both the source and the measurement apparatuses, as discussed by Morgan (2006) or Vervoort (2013). This appears to be for instance the strategy favoured by proponents of stochastic electrodynamics to explain nonlocality in quantum mechanics (de~la~Pe\~{n}a and Cetto 1996). We suspend judgement as to whether these cases should also count as `genuine' violations of MI, because it would appear that some mechanism of equilibration with the field might be needed to establish the dependence between the settings and the hidden variables, so that the distribution of hidden variables would not in general depend robustly on the settings. As a consequence, one might also be able to rule out such approaches experimentally because they would not be able to reproduce all violations of the Bell inequalities. In this sense, they would constitute merely another experimental loophole, and need not especially concern us.}

This paper is concerned exclusively with such genuine (systematic, robust, law-like) albeit putative violations of MI -- whether they be a matter of fundamental law or emerge only in certain regimes. This is reflected in our notation, where we have treated settings as parameters.\footnote{Much of the literature treats settings as random variables. This is clearly useful for some purposes, such as discussing common-cause explanations or quantifying the dependence between settings and hidden variables, but it also obscures the distinction between free and dependent variables. For this reason we find the parameter notation more appropriate to many modelling purposes in the setting of Bell's theorem, even irrespective of questions about MI. Treating settings as parameters can also be seen as more general, in the sense that it does not require the existence of a `big' probability distribution somehow describing both microscopic and macroscopic degrees of freedom (Butterfield 1992, Sect.~2).} We will identify conditions under which Alice can use such violations of MI to \emph{signal} to Bob at least `in principle', even if OI and PI are both satisfied. Deriving signalling from these conditions will not presuppose a hidden variables theory that reproduces all the predictions of quantum mechanics, but only models that reproduce well-established finite fragments of quantum mechanics. In this sense, we present an extension of what Shimony (1989) has called the programme of `experimental metaphysics', providing an analysis of quantum nonlocality that includes also violations of MI.

Two final remarks before we proceed. First, we shall usually say that Charlie `prepares an ensemble' of particle pairs and talk about `the ensemble' thus prepared. This is accurate in the sense that Charlie provides the pairs on which Alice and Bob will perform their measurements, and also in the sense that Charlie controls the quantum state of the pairs, which at least in part determines the distribution of hidden variables. But such terminology should not mislead one into neglecting that, if MI is violated, then there are in fact different probability distributions $\sigma_\rho^{IJ}(\lambda)$ compatible with the same preparation procedure performed by Charlie, and these distributions are a joint result of Alice's, Bob's and Charlie's choice of settings.\footnote{The same applies also to the case of `non-quantum' distributions, which we presently introduce, and to the case of `sub-ensembles', which we introduce in Section~\ref{strengthening}.} 

Second, one may adopt different views about causation: that it is an ontic notion or that it is an agent-centred notion, that it applies only at some fundamental level or that it applies at different levels of description. Our analysis is independent of such distinctions, since our notion of signalling (even of signalling `in principle') is operational.\footnote{This point will also be relevant for our discussion of Sen and Valentini (2020a,b) in Section~\ref{metaphysics}.} If one is happy with causation being an agent-centred high-level notion, our analysis is indeed about causation. If one is happy with level-dependent causation but not with agent-centred causation, we are only discussing special cases of causation (when it is used for signalling). If one is happy with neither, our analysis will simply be silent about it. In that case, we shall be drawing physical lessons about nonlocality but not metaphysical ones about causation...

%%%%%%%%%%%%%%%%%%%%%%%
%                                                               %
%%%%%%%%%%%%%%%%%%%%%%%
\section{Signalling in Principle}\label{signalling}
 We now make precise what we mean by `signalling in principle'. We wish to be able to define a signalling protocol, so the required notion needs to be implementable operationally: it needs to be a form of signalling that may be difficult to achieve in practice but is dynamically allowed. Our guide in this will be the notion of signalling in principle that applies to the case of violations of PI in de~Broglie--Bohm pilot-wave theory.

Take two spin-$\frac{1}{2}$ particles in a maximally entangled state. Let Alice perform a spin measurement on her particle. In pilot-wave theory, whether Alice's particle moves up or down in her magnetic field is determined by the particle's initial position, but -- just as in standard quantum mechanics -- how the two spin components of Alice's wave separate in the magnetic field (i.e.\ whether the up and down components move, respectively, up and down or the other way round) depends on the relation between the polarity and the sign of the field gradient. Therefore, given the initial position of her particle -- and even though Alice will not generally know what that position is -- whether the outcome of her spin measurement is up or down depends on how she chooses the polarity or gradient of her magnetic field (a choice of context that is immaterial for the outcome of her measurement in standard quantum mechanics). Furthermore, once the two components have separated and Alice's particle is trapped inside one of the two, the particle on \emph{Bob's} side is also guided only by one component of the wave: this is how the theory enforces the perfect (anti)\-correlations in an EPR experiment. But that means that Alice's choice affects not only the trajectory of her particle but also that of Bob's particle, and we have a clear case of action at a distance (see Figure~1).\footnote{For a lucid presentation, see Barrett (1999, Chap.~5).}
  \begin{figure}[h!]
    \centering
    \includegraphics[width=0.47\textwidth]{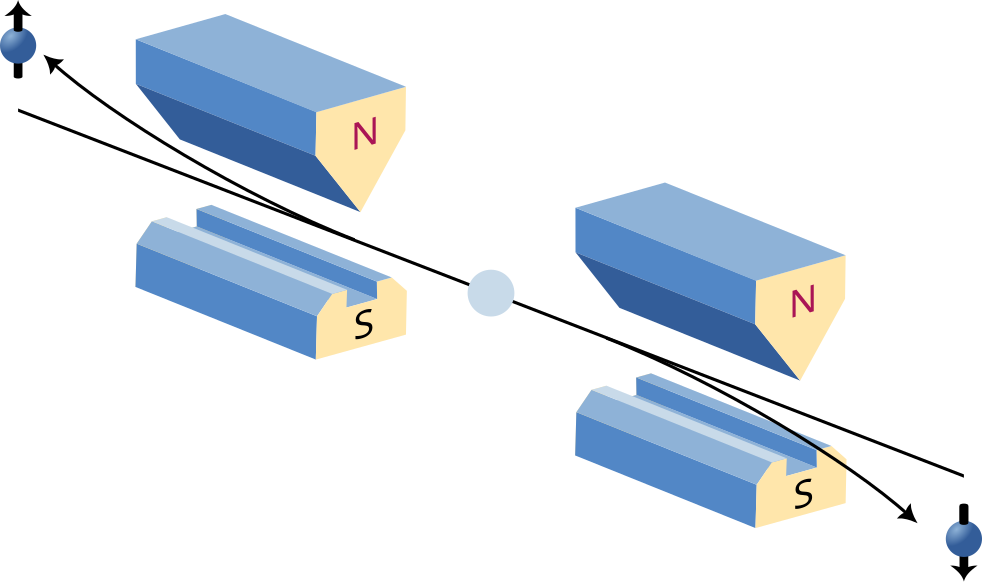}\qquad\includegraphics[width=0.45\textwidth]{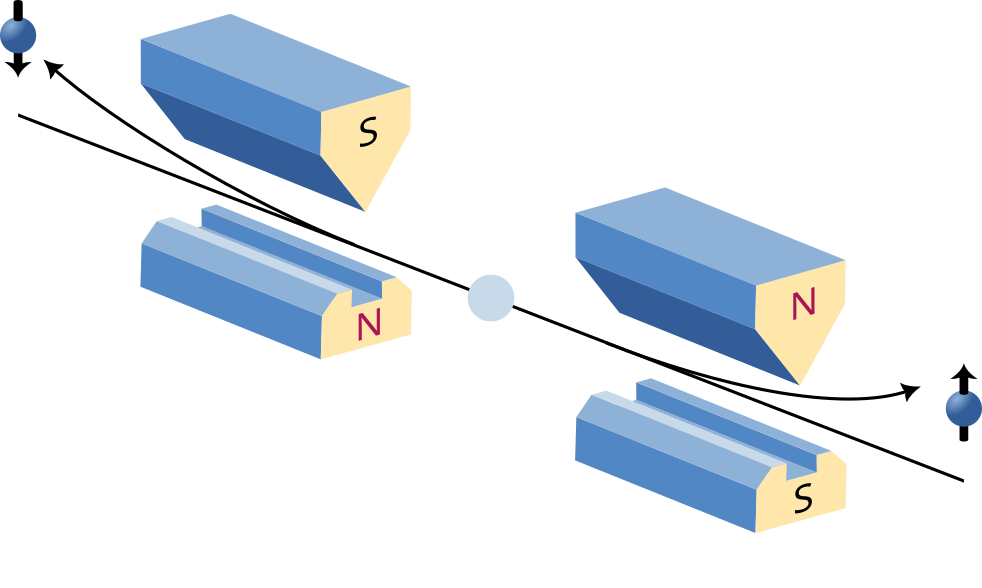}
    \caption{Action at a distance in pilot-wave theory: spin measurement with two choices of polarity for the field.}
  \end{figure}\\

One usually says that in this scenario there is `signalling in principle'. This is not to suggest one should analyse anthropocentrically the notion of action at a distance (which, as emphasised, occurs independently of whether Alice knows the position of her particle), rather it invites us to consider a thought experiment: if Alice in fact did know the initial position of her particle, she could use it to signal to Bob and thus \emph{reveal} the action at a distance inherent in the theory. This is not enough to explicate the specific sense of `in principle' that we require, however: it just shifts the question to whether and how `in principle' Alice  could know the initial position of her particle. There are two ways of filling in the details, which we argue are very different from an operational point of view.

One option is to take Alice to be a `sub-quantum demon', a being who is able to measure the position of a Bohmian particle \emph{without interacting quantum mechanically} with it (Valentini 1992, Chap.~7). Alice could then acquire the extra knowledge needed without otherwise altering the physical situation of the particles, thereby acquiring the ability to signal to Bob. This picture is intentionally analogous to that of Maxwell's demon, who also acquires knowledge of the state of motion of the molecules of a gas without physically interacting with them, thereby acquiring the ability to create disequilibrium without performing any work. It is also equally unrealistic, and can be `exorcised' by pointing out that if a physical being tries to measure a particle's position by interacting unitarily with it, they get entangled with the particle and thereby disturb its pilot wave, which in turn destroys their ability to signal.\footnote{The details are as follows. In order to be able to exploit the nonlocality of pilot-wave theory for signalling, the demon needs to know the position of the particle more precisely than according to the Born distribution. Suppose the demon attempts to gain more detailed information about the position of the particle by performing a realistic (i.e.\ quantum) measurement. As hinted above in the case of Alice's spin measurement, measurements in pilot-wave theory are modelled by applying the usual Hamiltonian interaction between the system and an appropriate `pointer' (in the Stern--Gerlach case, the position of the particle itself), resulting in entanglement. There is no collapse of the state: what happens instead is that the components of the state develop a negligible overlap in configuration space, specifically with respect to the coordinates of the pointer (the two components of the state are deflected upwards or downwards in the magnetic field) and will generally remain thus separated, so that the (total) system is `trapped' inside one of the  components, known as the `effective'  wave (or `full' wave -- as opposed to the other `empty' waves). Since the trajectory of a system is determined by the values of the pilot wave in a neighbourhood of the configuration of the system, only this component is relevant to guiding the further motion of the system, so there is an effective collapse from the point of view of how the state affects the motion of the particles, even though the evolution of the state itself remains unitary. In this sense a sub-quantum demon is unrealistic, because any realistic attempt at gaining information about the position of the particle will effectively collapse the wave. If the demon originally could merely estimate the position of the particle based on the Born probabilities defined by the original wave, now it can merely estimate the position of the particle based on the Born probabilities defined by the effective wave. And so the demon is back to square one. Similar `exorcisms' are well-known also for Maxwell's demon, if one treats it as a realistic physical system (Earman and Norton 1998, 1999).} Talking of sub-quantum demons as able to signal is a picturesque way of illustrating the notion of action at a distance in pilot-wave theory (just as Maxwell intended his own demon as a picturesque way of illustrating that the laws of thermodynamics do not have strict validity), but it does not operationalise `signalling in principle' in any useful way.

The second option is to consider `non-quantum' ensembles. Whatever protocol Alice chooses in order to try and signal using a shared ensemble of maximally entangled particles (i.e.\ whatever sequence of measurement contexts she chooses to adopt), she always has a small probability of success, because the Born distribution in pilot-wave theory is a (time-dependent) equilibrium distribution, and it is always possible that the ensemble she is using is an ensemble that is out of equilibrium.\footnote{Also in Maxwell's case, a real physical being might just be lucky enough that  particles with high velocity happen to move preferentially, say, from left to right whenever the trap door is opened.} What Alice needs in order to be able to signal with high probability is a \emph{reliable source} of such ensembles. Such sources would arguably be rare because interactions are known to typically induce relaxation to equilibrium, but -- crucially -- they are dynamically allowed in pilot-wave theory, yielding a realistic sense in which it is possible for Alice to signal `in principle'. Not only, if such reliable sources of non-quantum ensembles exist, they would also be operationally identifiable in advance, because randomly selected sub-ensembles would violate the predictions of quantum mechanics. A trusted source of non-quantum ensembles could then be used for signalling.\footnote{Note that a number of authors dismiss the notion of disequilibrium in pilot-wave theory and take it as law-like that initial conditions should be typical according to the Born measure (D\"{u}rr, Goldstein and Zangh\`{i} 1992). We need not argue for our differing position, since it is equally standard in the literature and is in fact what makes pilot-wave theory testable in principle. See Valentini (1991a,b; 1992, Chap.~3) and Valentini and Westman (2005) for classic discussions of disequilibrium and relaxation to equilibrium in pilot-wave theory.} 

When we presently construct a signalling protocol for theories violating MI, we shall accordingly disregard any `demonic' intuitions about Alice, Bob or Charlie having direct access or control over the hidden variables,\footnote{Indeed, such access would trivialise the claim of signalling in principle: if one of Alice, Bob or Charlie (who can subluminally inform the other two) could directly read off the distribution $\sigma_\rho^{IJ}(\lambda)$ (by observing the hidden variables in a sufficiently large randomly selected sub-ensemble), they would be able in general to infer back the values of $I$ and $J$, thus enabling at least one of Alice or Bob to signal.} and shall instead assume that the theory allows appropriate sources of non-quantum ensembles, i.e.\ ensembles where the distribution of hidden variables is \emph{not} given by $\sigma_\rho^{IJ}(\lambda)$ for some quantum state $\rho$. Whether this assumption is realistic will of course depend on the details of the theory under consideration (see also below, Sections~\ref{limitations} and \ref{metaphysics}).

In order to incorporate this assumption in our formal description, we need to include explicitly in our models an additional parameter $K$ describing Charlie's context of preparation. This is normally left implicit because one can imagine that $K$ is fixed in any particular experiment and, insofar as $K$ affects the distributions $\sigma_\rho^{IJ}(\lambda)$ via the quantum state $\rho$, it is already taken into account. Still, one can imagine that a distribution of hidden variables could explicitly depend on $K$ as $\sigma^{IJ}_K(\lambda)$ (and there is nothing particularly striking about this -- unlike the dependence on $I$ and $J$). As long as one assumes that $\sigma^{IJ}_K(\lambda)$ returns the expectation values $\mean{AB}_\rho^{IJ}$ for the quantities $A$ and $B$ (as defined in (\ref{2})), the details of $K$ do not matter (much like the polarity or gradient of a Stern--Gerlach magnet do not matter for the distribution of results in standard quantum theory). But if we consider also \emph{more general} distributions, then $K$ may come to play an essential role: it might label a source used by Charlie to reliably prepare ensembles that allow signalling between Alice and Bob.\footnote{In Section~\ref{strengthening} we shall use the notion of preparing a (non-quantum) \emph{sub-ensemble} of an ensemble $K$. Since Charlie is not a demon and cannot separate sub-ensembles depending on the values of the hidden variables, also this needs to be understood in terms of identifying reliable sources. See Section~\ref{strengthening} for details.} 

We shall thus generally understand a hidden variables model of a Bell experiment to include explicitly the possible dependence of the distributions on both Alice's and Bob's contexts of measurement and Charlie's context of preparation. For the sake of generality, we shall not distinguish between theories where there always exist contexts of preparation $K$ that correspond to quantum states $\rho$ and theories that do not. The definition of the expectation values (\ref{0b}), the assumption of MI (\ref{5}) and the Bell inequalities (\ref{6}) remain the same under the simple substitution of $\rho$ with $K$.\footnote{If one makes the parameter $K$ explicit, one might want to ask whether and how the expectation values $\mean{AB}_\lambda^{IJ}$ might also depend on $K$ (or whether perhaps $\lambda$ screens off $K$). We do not need to consider this for our purposes, because it will not affect the derivability of the Bell inequalities.}

%%%%%%%%%%%%%%%%%%%%%%%
%                                                               %
%%%%%%%%%%%%%%%%%%%%%%%
\section{Conditions for Signalling}\label{conditions}
We consider the following setup. Charlie prepares an ensemble of spin-$\frac{1}{2}$ pairs and Alice and Bob perform spin measurements. The hidden variables distributions have the form $\sigma^{IJ}_K(\lambda)$. If Charlie's choice of preparation context corresponds to a quantum state $\rho$, then clearly Alice and Bob cannot signal to each other, but we shall prove that if Charlie's choices of settings lead to certain kinds of non-quantum distributions, then Alice and Bob will in fact be able to signal. What is noteworthy is that our results hold for any model of bipartite spin experiments \emph{irrespective} of whether MI is satisfied: they will tell us nothing new if signalling is due to violations of PI, but they will apply also to models that satisfy PI (and OI), where signalling must then be due to violations of MI. 

The crucial ingredient we shall need for our results is the following `equiprobability theorem'.\footnote{This theorem was first proved by Squires, Hardy and Brown (1994) for the special case of deterministic hidden variables theories. The version we use here is closest to that of Barrett, Kent and Pironio (2006). It is also a crucial step in the proof of the Colbeck--Renner theorem, on which more below. The name 'equiprobability theorem' we believe is due to Leifer (2014).}
  \begin{quote}
     {\bf Theorem~0}:\\
     Take a hidden variables model of spin measurements on pairs of 
     spin-$\frac{1}{2}$ particles satisfying PI and MI. 
     If it reproduces the conditional probabilities between arbitrarily close spin directions of a 
     maximally entangled state, then for almost all individual $\lambda$ the 	
     results of any spin measurement on either particle are equiprobable.
  \end{quote}
Note that equiprobability of measurement results for given $\lambda$ is trivially equivalent to $\mean{A}^{IJ}_\lambda=\mean{B}^{IJ}_\lambda=0$, which by PI is further equivalent to $\mean{A}^I_\lambda=\mean{B}^J_\lambda=0$.  
  \begin{figure}[h!]
    \centering
    \includegraphics{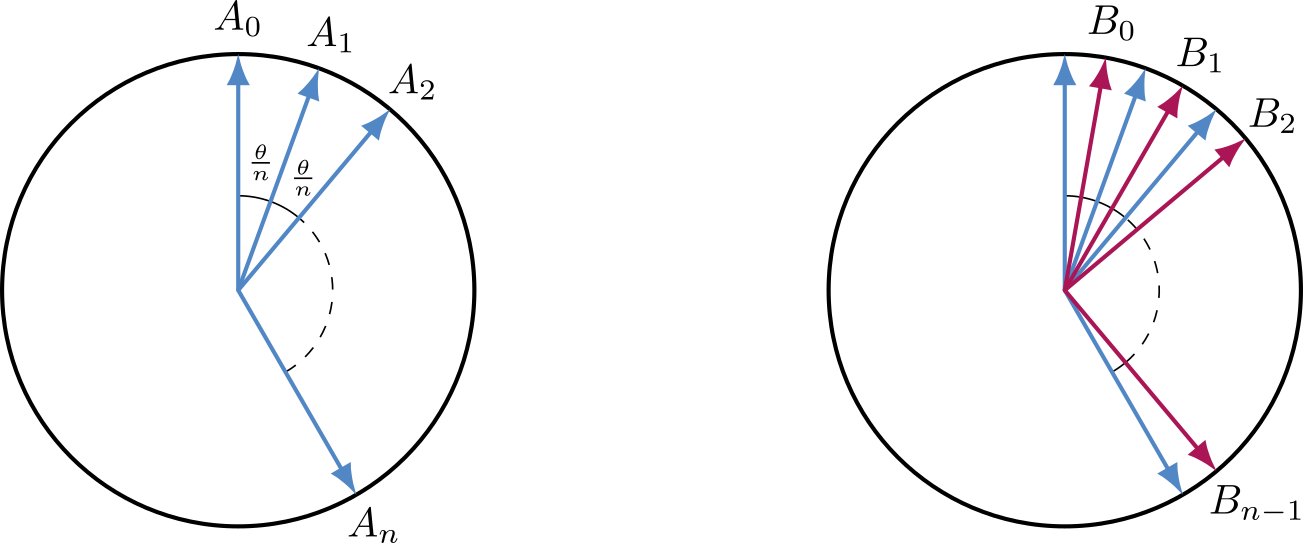}
    \caption{Spin directions for the proof of the equiprobability theorem.}
  \end{figure}

The main idea behind the proof is that, considering a chain of coplanar spin measurements $A_i$ and $B_j$ alternating between Alice's and Bob's side (see Figure~2), on average (where averaging requires MI) if the correlations along the chain (where each step requires PI) deviate from perfect (anti)\-correlations by a bounded amount, then the difference in probability for the spins connected by the chain (in particular two opposite spin vectors $\ket{A=+1}$ and $\ket{A=-1}$) will also be bounded accordingly. This is important enough to be stated as a separate theorem.
  \begin{quote}
     {\bf Theorem~0$'$}:\\
     Take a hidden variables model of spin measurements on pairs of 
     spin-$\frac{
1}{2}$ particles satisfying PI and MI. Consider a chain of $2n$ 
     pairs of coplanar spin directions at 
     angles $\frac{\pi}{2n}$ (measured alternating between Alice's and Bob's side) that connect a pair of vectors $\ket{A=+1}$ and $\ket{A=-1}$. If Charlie prepares an 
     ensemble $K$ such that, for some small $\delta>0$ and on average for all links in the chain, either
       \begin{equation}
         \mean{A_iB_j}_K^{I_iJ_j}\geq1-2\delta
         \label{corr}
       \end{equation}  
(high correlations) or 
       \begin{equation}
         -\mean{A_iB_j}_K^{I_iJ_j}\geq1-2\delta
         \label{anticorr}
       \end{equation}  
(high anticorrelations), then the corresponding 
     marginal deviates from equiprobability at most by
     \begin{equation}
         \big|\!\mean{A}_K^{IJ}\!\big|=\big|\!\mean{A}_K^{I}\!\big|\leq2n\delta
         \label{8}
       \end{equation}
     (or similarly on Bob's side). 
  \end{quote}
 Theorem~0 then follows by taking a maximally entangled state, for which $\delta$ is of order $\frac{1}{n^2}$, and letting $n\rightarrow\infty$: see Appendix~\ref{A} for details. Note also that Theorem~0$'$ is \emph{robust} under small deviations in the given spin directions.
 
Theorems~0 and 0$'$ show that if equiprobability fails at the hidden level, then PI or MI must fail. We shall now prove our first main result, which extends these theorems by showing that if a failure of equiprobability manifests itself at the \emph{operational} level, then this violation of PI or MI leads to \emph{signalling}. 
  \begin{quote}
     {\bf Theorem 1}:\\
     Let Charlie prepare an ensemble $K$ of spin-$\frac{1}{2}$ pairs and let them 
     send it to 
     Alice and Bob. If the ensemble reproduces the conditional probabilities between arbitrarily 
     close spin directions of a maximally entangled state, but for some values of $I$ and $J$
       \begin{equation}
          \mean{A}_K^{IJ}\neq 0 
          \label{8b}
       \end{equation}   
     (or $\mean{B}_K^{IJ}\neq 0$), then at least one of Alice or Bob can signal.
  \end{quote}
  \begin{quote}
     {\bf Theorem 1$'$}:\\
     Let Charlie prepare an ensemble $K$ of spin-$\frac{1}{2}$ pairs and let them send it to 
     Alice and Bob. Consider a chain of $2n$ 
     pairs of coplanar spin directions at 
     angles $\frac{\pi}{2n}$ (measured alternating between Alice's and Bob's side) that connect a pair of vectors $\ket{A=+1}$ and $\ket{A=-1}$. If for some small $\delta>0$ and on average for all links in the chain the ensemble satisfies either (\ref{corr}) or (\ref{anticorr}), but for some values of $I,J$ the corresponding marginal violates 
     equiprobability by
       \begin{equation}
         \big|\!\mean{A}_K^{IJ}\!\big|>2n\delta
         \label{9}
       \end{equation}
      (or similarly on Bob's side), then at least one of Alice or Bob can signal.
  \end{quote}
We give direct proofs in Appendix~\ref{B}, but Theorems~1 and 1$'$ can also be seen to be consequences of Theorems~0 and 0$'$, as follows. Alice, Bob and Charlie control their experimental settings $I$, $J$ and $K$, respectively. We assume Alice and Bob cannot signal to each other, and define a trivial hidden variables model in which the only hidden variable is $K$. (This is analogous to taking quantum theory as a trivial hidden variables theory with the quantum state as the hidden variable.) Since Charlie has complete control over the hidden variable, this model satisfies MI whether or not it is satisfied in any more detailed model. And since by construction Alice and Bob can know the value of $K$, the model must also satisfy PI otherwise at least one of them could signal.\footnote{One might want to say that Alice and Bob are \emph{actual} demons for this model!} We can thus apply Theorems~0 and 0$'$ to our trivial hidden variables model, and -- depending on the correlations in the ensemble -- the results of Alice's and Bob's local measurements must be equiprobable,  or they must deviate from equiprobability at most by (\ref{8}). 

A few remarks are in order. If we assume a hidden variables model that satisfies MI, we have learned nothing new: failure of equiprobability at the operational level requires its failure at the hidden level, but then Theorems~0 and 0$'$ show that the hidden variables model must violate PI, and Valentini (2002) has shown that any non-quantum distribution in a hidden variables theory that violates PI and reproduces perfect (anti)correlations can be used to signal.\footnote{To be precise, Valentini assumes that the hidden variables theory is deterministic, but the proof goes through also if one assumes merely that the theory reproduces perfect (anti)correlations.} If instead we assume a hidden variables model that satisfies PI, then Theorems~0 and 0$'$ show that it violates MI, and our Theorems~1 and 1$'$ show that this violation can \emph{also} be used to signal.

It is worth emphasising an aspect of the proof just given, which at first sight looks like a `trick': at a coarser level one might have a violation of PI, but at a finer level one has violation of MI. A little reflection, however, shows that this is not surprising. In our trivial model, we have
  \begin{equation}
    \mean{A}_K^{IJ}=
    \int\sigma_K^{IJ}(\lambda)d\lambda\,\mean{A}_\lambda^{IJ}\ .
    \label{Kprobs}
  \end{equation}
It is clear that a violation of PI at the coarser level may derive from a violation of PI \emph{or MI} at the finer level (and that at intermediate levels of detail the possible violation of MI may be partially absorbed into violation of PI). Far from being a trick, this is key to understanding our results.\footnote{See Leegwater (2016, Sect.~9) for a previous use of a similar trick.} Because we focus on the operational level (left-hand side of (\ref{Kprobs})), our results are neutral with respect to any underlying mechanisms (right-hand side of (\ref{Kprobs})), be they action at distance, retrocausation, a combination of both, or what not. As a matter of fact, such level-relativity of explanatory mechanisms is not really novel. We know very well that a model may violate OI at a coarser level while it violates PI at a finer level (think of quantum mechanics and pilot-wave theory), and proponents of stochastic electrodynamics argue both that at a coarse-grained level the theory reproduces Nelson's stochastic mechanics (which violates PI), but that stochastic electrodynamics itself merely violates MI.\footnote{See de~la~Pe\~{n}a, Cetto and Vald\'{e}s-Hern\'{a}ndez (2020) for a proposed reduction of stochastic mechanics to stochastic electrodynamics, and footnote 10 above for the possible connection between stochastic electrodynamics and violations of MI.}

Finally, we note that on previous occasions we have given talks presenting a version of our results as an extension of the Colbeck--Renner theorem.\footnote{See Colbeck and Renner (2011, 2016), Leegwater (2016), Hermens (2020).} This theorem (in the formulation of Leegwater (2016)) states that any hidden variables theory that satisfies MI will violate PI if it is non-trivial (in the sense that it makes non-quantum predictions for a set of hidden variables not of measure zero), and the proof of the theorem includes the equiprobability theorem as a crucial step. The same trick of applying the Colbeck--Renner theorem to the trivial hidden variables model yields a similar extension also of that theorem:  \emph{without} assuming MI, any theory allowing non-quantum ensembles at the operational level will allow for signalling. Focusing on  equiprobability has the advantage that our results can be made quantitative, and in the next section we shall use this fact to connect them to violations of the Bell inequalities.\footnote{To our knowledge, the first explicit proof of a version of Theorem~0$'$, in fact motivated by the Colbeck--Renner theorem, is by Stuart \emph{et al.}\ (2012, Appendix, Lemma~1), who propose and report on an experiment to constrain possible violations of equiprobability at the hidden level (i.e.\ to constrain the allowable non-triviality of hidden variables theories). Two of us (GB and GL) are fans of Colbeck--Renner, but the third (RH) convinced us it was wiser to focus on equiprobability, because it rests on fewer assumptions.}

%%%%%%%%%%%%%%%%%%%%%%%
%                                                               %
%%%%%%%%%%%%%%%%%%%%%%%
\section{Extending Bell's Theorem}\label{strengthening}
We have identified conditions under which a certain form of non-triviality in hidden variables models leads to signalling. In this section we shall point out that violation of the Bell inequalities requires this form of non-triviality, so that under the same conditions it leads to signalling. We thus effectively show that imposing no-signalling allows one to derive the Bell inequalities. The strength of our results lies in the fact that they are neutral as to the mechanism that underlies such signalling, whether it be violations of PI or of MI, or both. 

As such, our results are different from existing derivations of the Bell inequalities that do not impose MI (briefly mentioned in Section~\ref{intro}). These other derivations focus exclusively on MI, limiting the strength of the correlations between the hidden variables and Alice's and Bob's settings (and require treating settings as random variables). 

We start by introducing a useful distinction. Take an ensemble $K$. By a \emph{sub-distribution} of a distribution $\sigma^{IJ}_K(\lambda)$ we simply mean the distribution over the hidden variables in any theoretically definable sub-ensemble of $K$, irrespective of whether this can be operationally prepared or not. We shall reserve the term \emph{sub-ensemble}, however, for one that can in fact be prepared operationally. This needs to be spelled out with some care. As mentioned towards the end of Section~\ref{neglected}, every context of preparation $K$ labelling a procedure used by Charlie in fact corresponds to several probability distributions $\sigma_K^{IJ}(\lambda), \sigma_K^{I'J}(\lambda),\sigma_K^{IJ'}(\lambda),\sigma_K^{I'J'}(\lambda)$ depending on the contexts of measurement chosen by Alice and Bob. In the case of models satisfying MI, an ensemble is characterised just by a single probability distribution $\sigma_K(\lambda)$, and one defines mixtures of any two ensembles $\sigma_{K_1}(\lambda)$ and $\sigma_{K_2}(\lambda)$ with the weights $0\leq\alpha,\ 1-\alpha\leq1$ via
  \begin{equation}
    \sigma_K(\lambda)=\alpha\sigma_{K_1}(\lambda)+(1-\alpha)\sigma_{K_2}(\lambda) \ .
    \label{mix1}
  \end{equation}
If instead MI is not satisfied, we can define a mixture analogously as an ensemble characterised \emph{for all $I,J$} by the distributions
  \begin{equation}
    \sigma_K^{IJ}(\lambda)=\alpha\sigma_{K_1}^{IJ}(\lambda)+
    (1-\alpha)\sigma_{K_2}^{IJ}(\lambda) \ . 
    \label{mix2}
  \end{equation} 
In the case of (\ref{mix1}), the existence of a procedure that prepares the given mixture is trivially guaranteed: mix together ensembles prepared via $K_1$ and $K_2$ in the proportion determined by a flip of an appropriately weighted coin. In the case of (\ref{mix2}), it is non-trivial (if presumably natural) to assume that this same mixing procedure works uniformly across all of Alice's and Bob's contexts of measurement $I,J$.\footnote{The worry is that, just as different settings $I',J'$ for Alice's and Bob's measurements in the future of the preparation $K_1$ lead to different sub-distributions $\sigma^{I'J'}_{K_1}$, so another such procedure, namely mixing with the ensemble $K_2$, might also lead to the corresponding sub-distribution in the resulting ensemble being different from $\sigma^{IJ}_{K_1}$. In this case, the natural operational procedures for mixing two ensembles might not produce the mathematical mixture (\ref{mix2}). Thus, the property that mixtures are in fact operationally well-defined will in general need to be established explicitly in any concrete theory that violates MI.} If we do assume that mixtures are operationally well-defined, a \emph{sub-ensemble} $K_1$ of an ensemble $K$ can then be defined as an ensemble that operationally mixed with another suitable ensemble $K_2$ yields back ensemble $K$. 

We shall now prove two lemmas (for which we only need the notion of a sub-distribution). The first yields a Bell-like inequality without assuming MI -- or even PI -- from a constraint of the form (\ref{8}) that limits how much a model deviates from equiprobability.
  \begin{quote}
     {\bf Lemma 1}:\\
     Let Charlie prepare an ensemble $K$ of pairs of spin-$\frac{1}{2}$ particles, with Alice 
     and Bob measuring spins $A,A',B,B'$ in the contexts $I,I',J,J'$, respectively. Assume a 
     hidden variables model of this setup that satisfies OI.  
     
     If
           \begin{equation}
    \big|\mean{AB}_K^{IJ} + 
    \mean{A'B}_K^{I'J} + 
    \mean{AB'}_K^{IJ'} - 
    \mean{A'B'}_K^{I'J'}\big|> 2\varepsilon + 2     
    \label{10b}
  \end{equation}
     for any $1>\varepsilon\geq0$ (or similarly with the minus sign positioned differently), then there is at least one combination of 
     Alice's and Bob's settings, say $IJ$, and 
     there is a sub-distribution $\sigma_0^{IJ}(\lambda)$ with weight 
     $\frac{1}{2}\leq\alpha\leq 1$ in the corresponding $\sigma_{K}^{IJ}(\lambda)$, such that
       \begin{equation}
         \big|\mean{A}^{IJ}_0\big|>\varepsilon 
         \label{10}
       \end{equation}   
       (and similarly on Bob's side\footnote{That is, there are a (possibly different) combination of 
       settings and a (possibly different) sub-distribution such that also a marginal on Bob's side 
       deviates from equiprobability by $\varepsilon$.}). 
  \end{quote}
The proof is  straightforward but uninteresting (see Appendix~\ref{C} for details). 
The interesting point is that condition (\ref{10}) in the conclusion of Lemma~1 has the same form as condition (\ref{9}) in the premises of Theorem~1$'$. As regards the other premise of Theorem~1$'$, that the model should exhibit high (anti)\-correlations  along a chain of spin pairs, it is trivial to prove the following (see again Appendix~\ref{C}).
  \begin{quote}
    {\bf Lemma 2}:\\
    Let Charlie prepare an ensemble $K$ of pairs of spin-$\frac{1}{2}$ particles, and let 
    Alice and Bob measure spins $A,B$ in the contexts $I,J$. Take any sub-distribution 
    $\sigma_1^{IJ}(\lambda)$ with weight $0<\alpha\leq 1$ (and \emph{a fortiori} with weight $\tfrac{1}{2}\leq\alpha\leq 1$)
    in $\sigma_{K}^{IJ}(\lambda)$. Then 
      \begin{equation}
        |\mean{A_iB_j}_K^{I_iJ_j}|\geq1-2\gamma\;\Longrightarrow\;
        |\mean{A_iB_j}_1^{I_iJ_j}|\geq1-\frac{2\gamma}{\alpha} \ .
      \end{equation}
  \end{quote}

Given the results of the previous section, the upshot of Lemmas~1 and 2 is intuitively that, assuming a hidden variables model satisfying OI, the experimental violation of the Bell inequalities \emph{guarantees the existence of sub-ensembles that allow for signalling} if only these can in fact be prepared. 

We are now ready to state the main results of this section.
\begin{quote}
     {\bf Theorem 2}:\\
     Take an ensemble $K$ of pairs of spin-$\frac{1}{2}$ particles, and let 
     Alice and Bob measure spins $A,A',B,B'$, as well as four chains of 
     $2n$ pairs of coplanar spin directions at angles $\frac{\pi}{2n}$ 
     (alternating 
     between Alice's and Bob's side) that connect, respectively, $\ket{A=+1}$ to $\ket{A=-1}$, 
     $\ket{A'=+1}$ to $\ket{A'=-1}$, $\ket{B=+1}$ to $\ket{B=-1}$, and $\ket{B'=+1}$ to $\ket{B'=-1}$. Assume a hidden 
     variables model of this setup that satifies OI, assume that for each of the chains and on average for all the links in the chain either 
       \begin{equation}
         \mean{A_iB_j}_K^{I_iJ_j}\geq1-2\gamma
         \label{gamma1}
       \end{equation}  
or 
       \begin{equation}
         -\mean{A_iB_j}_K^{I_iJ_j}\geq1-2\gamma
         \label{gamma2} 
       \end{equation}
for some $\gamma>0$, and assume 
           \begin{equation}
    \big|\mean{AB}_K^{IJ} + 
    \mean{A'B}_K^{I'J} + 
    \mean{AB'}_K^{IJ'} - 
    \mean{A'B'}_K^{I'J'}\big|> 2\varepsilon + 2     
   \label{eq23}
  \end{equation}
     (or similarly with the minus sign positioned differently) for some $\varepsilon\geq4n\gamma$.       
    
    If Charlie can prepare appropriate sub-ensembles of $K$, then at least one of Alice 
    or Bob can signal.  
  \end{quote} 
  \begin{quote}
    \emph{Proof}.\\
    By Lemma~2, any sub-ensemble with weight $\tfrac{1}{2}\leq\alpha\leq1$ in $K$ satisfies for each chain on average either (\ref{corr}) or (\ref{anticorr}) with $\delta=2\gamma$. By 
    Lemma~1, one of these sub-ensembles also satisfies (\ref{10}) for some combination of settings. Since $\varepsilon\geq4n\gamma$, this sub-ensemble satisfies 
    (\ref{9}) for that combination of settings, and thus both the premises of Theorem~1$'$. Therefore, if Charlie can prepare such a sub-ensemble of $K$, then at least one of 
    Alice or Bob can signal.   
  \end{quote}

As before, if we assume MI we have learnt nothing new: the standard version of Bell's theorem implies that, given OI, if Charlie is able to prepare appropriate sub-ensembles of an ensemble that violates the Bell inequalities, then one of Alice or Bob can signal by exploiting the violation of PI. But if MI is violated, then we have an \emph{extension} of Bell's theorem: under the assumption that an ensemble reproduces an additional finite set of conditional probabilities, and given OI, if Charlie is able to prepare appropriate sub-ensembles of an ensemble that violates the Bell inequalities, we have shown that \emph{also in this case} one of Alice or Bob can signal; and if PI is satisfied they can only do so \emph{by exploiting the violation of MI}.\footnote{We cannot relax the assumption of OI, because in that case quantum mechanics itself provides a counterexample.} 

For $n=6$ the angle between each spin direction in a chain $A_0,B_0,A_1,B_1,\ldots A_6$ is $\tfrac{\pi}{12}$, thus the successive angles between $A_0$, $B_1$, $A_3$ and $B_4$ are equal to $\tfrac{\pi}{4}$. These form a set of directions that for a maximally entangled state yield the maximum quantum violation of the Bell inequalities. Using these directions (and taking into account that a measurement of any $A_i$ or $B_j$ also doubles up as a measurement of the opposite spin direction -- in particular $A_0$ and $A_6$), this single chain contains all the measurements we need in the premises, and we get a particularly simple special case of Theorem~2. 
 \begin{quote}
     {\bf Theorem 2$'$}:\\
     Take an ensemble $K$ of pairs of spin-$\frac{1}{2}$ particles, and a sequence 
     of 12 coplanar spin directions $A_0,\ldots,A_5$, $B_0,\ldots,B_5$ (alternating 
     between 
     Alice's and Bob's side) at successive angles $\frac{\pi}{12}$. Let Alice and Bob 
     measure along the 16 pairs of directions $(A_0,B_0), (B_0,A_1), (A_1,B_1),
     (B_1,A_2),\ldots, (A_5,B_5)$ and $(B_5,A_0)$, as well as $(A_0,B_1),
     (A_0,B_4),(A_3,B_1)$ and $(A_3,B_4)$. Assume a hidden variables model of 
     this setup that satifies OI, and assume that on average for the 12 measurements 
     along the chain either (\ref{gamma1}) or (\ref{gamma2}) for some 
     $\gamma>0$ and that the last 4 measurements satisfy (\ref{eq23}) for some 
     $\varepsilon\geq24\gamma$.
     
    If Charlie can prepare appropriate sub-ensembles of $K$, then at least one of Alice 
    or Bob can signal.  
  \end{quote}
If $K$ is a maximally entangled quantum state, we theoretically
expect $\varepsilon=\tfrac{\sqrt{2}-2}{2}\approx0.414$, and the corresponding correlations along the chain to be given by $1-2\sin^2(\tfrac{\pi}{24})$, i.e.\ $\gamma\approx0.017$ and $4n\gamma\approx 0.408$. Larger values of $n$ in Theorem~2 will provide increased robustness under variations both in the preparation of the ensemble and in the spin directions measured.

%%%%%%%%%%%%%%%%%%%%%%%
%                                                               %
%%%%%%%%%%%%%%%%%%%%%%%
\section{The Schulman Model}\label{limitations}
Up to this point, our discussion has been rather abstract. It is time to introduce a concrete example of a hidden variables model that violates MI and that may be a candidate of a model that allows for signalling. 

Consider first a toy model of bipartite spin measurements. Let the possible values for $\lambda$ be pairs of spin values, and for any $I, J$ let the distribution $\sigma_\rho^{IJ}(\lambda)$ resulting from Charlie's preparation of the quantum state $\rho$ be concentrated on values $(A=\pm1,B=\pm1)$ of the corresponding $A, B$ and be equal to
  \begin{equation}
    \sigma_\rho^{IJ}(A=\pm1,B=\pm1)=
    \mbox{Tr}(\rho \ket{A=\pm1}\!\bra{A=\pm1}\otimes\ket{B=\pm1}\!\bra{B=\pm1})
    \label{12}
  \end{equation}
(i.e.\ the corresponding quantum probabilities). 

This toy model clearly reproduces the quantum predictions for Bell experiments by violating MI. But it is not detailed enough to allow us to say whether distributions other than (\ref{12}) that would lead to signalling might also be possible, let alone might be reliably prepared. In order to do so, we need to specify further what determines the dependence between the source and Alice's and Bob's settings. 

A significant step in this direction is provided by a model introduced by Wharton (2014). He calls it the `Schulman model', because it is a generalisation to two maximally entangled particles of a single-particle one proposed by Schulman (2012).\footnote{We largely follow the presentation in Almada \emph{et al.}\ (2016), which is more detailed than the original one by Wharton.} As we argue below, this model by Wharton not only uses violations of MI to reproduce the quantum predictions for bipartite spin measurements, but allows for ensembles that lead to signalling in the sense of our theorems above. 

Schulman considers a single spin-$\frac{1}{2}$ particle subject to two successive measurements along directions $A$ and then $B$. His is not a hidden variables model but a proposal for a smooth evolution of the state vector of the particle between two measurements. Taking the outcomes of the measurements as boundary conditions for this evolution, in general the particle will have to undergo some anomalous rotation on the Bloch sphere. Also taking into account the macroscopic asymmetry in the boundary conditions, where the outcome of the $A$-measurement is pre-selected before the particle is subject to the $B$-measurement (say, $A=+1$ and either $B=+1$ or $B=-1$), there will be two possible anomalous rotations by a net angle $\theta\ (\!\!\!\!\mod 2\pi)$ or $\pi+\theta\ (\!\!\!\!\mod 2\pi)$ (where $\theta$ is the angle between $A$ and $B$). We can now ask for the relative probabilities of these two anomalous rotations. Schulman's \emph{ansatz} is that between the two measurements the particle will undergo a rotation with net angle $\alpha$ with probability proportional to
  \begin{equation}
    w(\alpha)\ldef\frac{1}{\alpha^2+\gamma^2} \ ,
    \label{Sch1}
  \end{equation}
where $\gamma>0$ is some very small parameter. Given that the same angle $\alpha$ will result from rotations by $2\pi n+\alpha$ for arbitrary $n$, Schulman calculates that the probabilities for the two outcomes stand in the ratio
  \begin{equation}
    \frac{p(\theta)}{p(\pi+\theta)}
    =\frac{\sum_{n=-\infty}^\infty w(2\pi n+\theta)}{\sum_{n=-\infty}^\infty w(2\pi n+
       \pi+\theta)}
    =\frac{\cos^2(\tfrac{\theta}{2})+\sin^2(\tfrac{\theta}{2})\tanh^2(\tfrac{\gamma}{2})}{\sin^2(\tfrac{\theta}{2})+
       \cos^2(\tfrac{\theta}{2})\tanh^2(\tfrac{\gamma}{2})} \ .
       \label{Sch2}   
  \end{equation}
Since $\gamma$ is assumed to be very small, (\ref{Sch2}) is very close to $\frac{\cos^2(\theta/2)}{\sin^2(\theta/2)}$, so that (to an arbitrary approximation and once probabilities are normalised)
  \begin{equation}
    p(\theta)\approx\cos^2(\tfrac{\theta}{2})\qquad\mbox{and}\qquad 
    p(\pi+\theta)\approx\sin^2(\tfrac{\theta}{2}) \ .
    \label{Sch3}
  \end{equation}  
We thus have a model for an anomalous evolution of the quantum state of a single particle between successive spin measurements that reproduces the Born rule. 
This model is often described as retrocausal because the choice of the second measurement direction influences the probabilities for the anomalous rotation (if any) between the two measurements. But this reading is not forced upon us, because one could also imagine that all the anomalous rotation happens during the interaction with the second measuring apparatus.

Wharton's two-particle model by contrast is an explicit hidden variables model that provides a retrocausal mechanism violating MI. When Charlie prepares a particle pair in a maximally entangled state, each particle is assumed to be also described by a hidden spin vector in some direction, which after possibly undergoing some anomalous rotation is revealed as Alice's or Bob's measurement outcome. And, as we shall see, this spin vector generally depends on Alice's or Bob's measurement settings.

Wharton's model is based on two simple assumptions: (a) that the hidden spins of the pairs are constrained at preparation to be all parallel (or all anti-parallel); (b) that the unnormalised joint probability for anomalous rotations by the net angles $\alpha$ for Alice's particle and $\beta$ for Bob's particle is 
  \begin{equation}
    w(\alpha)w(\beta)
    \label{Sch3a}
  \end{equation} 
  (with $w(\cdot)$ as in (\ref{Sch1})). He then applies Schulman's \emph{ansatz} to both particles \emph{separately}. 

Crucially, unlike in the single-particle case, the initial spins of the particles are not fixed by the boundary conditions, so we are interested in the relative probabilities of pairs of rotations which combined with each other yield the angle $\theta$ or $\pi+\theta$. Wharton notes that (\ref{Sch1}) implies $w(0)\gg w(\alpha)$ for $\alpha\gg\gamma$. And since $\gamma$ is very small, this means that $w(0)\gg w(\alpha)$ for all appreciable $\alpha$. But this in turn means that both 
  \begin{equation}
    w(0)w(\beta)\gg w(\alpha)w(\beta)
    \label{Sch4}
  \end{equation} 
and 
  \begin{equation}
    w(\alpha)w(0)\gg w(\alpha)w(\beta)
    \label{Sch5}
  \end{equation} 
for all appreciable $\alpha$ and $\beta$. Therefore, of all possible combinations of anomalous rotations for the two particles that can lead to outcomes along $A$ and $B$, the overwhelmingly most probable ones are either such that only Bob's hidden spin undergoes rotation, so that both hidden spins are initially (anti-)parallel to $A$, or such that only Alice's hidden spin undergoes rotation, so that both spins are initially (anti-)parallel to $B$. This is a retrocausal model that violates MI because Alice's and Bob's choices influence not only the probabilities for the anomalous rotations of the hidden states of the particles between preparation and measurements (which one may imagine happening only during the interaction of the particles with Alice's and Bob's apparatuses), but also on which hidden states the initial distribution is concentrated.

The predicted probabilities for the outcomes now are as follows. For definiteness, let us take the case in which Charlie prepares the pairs such that the hidden spins of the particles in each pair are parallel to each other. There are four possible initial pairs of hidden spins (neglecting ones with very small probability):
  \begin{equation}
      (A=+1,A=+1),\quad(B=+1,B=+1),\quad(A=-1,A=-1)
      \quad\mbox{and}\quad(B=-1,B=-1)\ .
    \label{Sch6}
  \end{equation}
With overwhelming probability, the initial state $(A=+1,A=+1)$ produces the outcome $A=+1$ and the outcomes $B=+1$ and $B=-1$ in the ratio $\frac{\cos^2(\theta/2)}{\sin^2(\theta/2)}$, and similarly for the other initial states. Therefore, setting
  \begin{equation}
  \begin{split}
    \alpha &\ldef\sigma_\rho^{IJ}(A=+1,A=+1)\ ,\\ 
    \beta &\ldef\sigma_\rho^{IJ}(B=+1,B=+1)\ ,\\ 
    \gamma &\ldef\sigma_\rho^{IJ}(A=-1,A=-1)\ ,\\ 
    \delta &\ldef\sigma_\rho^{IJ}(B=-1,B=-1)\ ,
    \label{Sch7}
  \end{split}
  \end{equation}
we have (to an arbitrary degree of approximation)
  \begin{equation}
  \begin{split}
    & p_\rho^{IJ}(A=+1,B=+1) \approx(\alpha+\delta)\cos^2(\tfrac{\theta}{2})\ ,\\
    & p_\rho^{IJ}(A=+1,B=-1) \approx(\alpha+\beta)\sin^2(\tfrac{\theta}{2})\ ,\\
    & p_\rho^{IJ}(A=-1,B=-1) \approx(\beta+\gamma)\cos^2(\tfrac{\theta}{2})\ ,\\
    & p_\rho^{IJ}(A=-1,B=+1) \approx(\gamma+\delta)\sin^2(\tfrac{\theta}{2})\ .
    \label{Sch8}
  \end{split}  
  \end{equation}
Note that since $\alpha+\beta+\gamma+\delta=1$, we have $p_\rho^{IJ}(A=B) \approx\cos^2(\tfrac{\theta}{2})$ and 
$p_\rho^{IJ}(A\neq B) \approx\sin^2(\tfrac{\theta}{2})$, i.e.\ (\ref{Sch8}) reproduces arbitrarily closely all the correlations of a maximally entangled state with perfect correlations (and, analogously, for the case of anti-parallel hidden spins it reproduces the correlations of a state with perfect anti-correlations). If one additionally assumes, as Wharton tacitly does,\footnote{Email from Ken Wharton to GB, 24 December 2022.} that histories with the same amount of anomalous rotation have the same probability, then $\alpha=\beta=\gamma=\delta=\frac{1}{4}$ and (\ref{Sch8}) reproduces arbitrarily closely also the marginals predicted by the Born rule. More generally, the Born rule is recovered iff 
  \begin{equation}
    \alpha+\delta\;=\;\alpha+\beta\;=\;\beta+\gamma\;=\;\gamma+\delta\;=\;\tfrac{1}{2} \ ,
  \end{equation}
 i.e.\ iff
  \begin{equation}
     \alpha=\gamma\ ,\quad\beta=\delta\ ,
     \quad\mbox{and}\quad\alpha+\delta=\tfrac{1}{2}\ . 
     \label{Sch9}
  \end{equation}

We see that Schulman and Wharton's \emph{ansatz} fixes the domain where $\sigma_\rho^{IJ}(\lambda)$ is non-negligible as well as the correlations in the ensemble, but underdetermines the full distribution $\sigma_\rho^{IJ}(\lambda)$.\footnote{There may be some dynamical justification for Wharton's additional assumption and thus for privileging as `equilibrium'  the uniform distribution over the hidden spin pairs (or more generally a distribution of the form (\ref{Sch9}) leading to the Born rule). But if we take seriously the lesson of statistical mechanics or of pilot-wave theory, such considerations at most provide naturalness arguments for equilibrium, and other distributions remain dynamically possible.} It is perhaps not far-fetched to imagine that while Alice's and Bob's choices determine the possible hidden spin states, the coefficients $\alpha, \beta,\gamma$ and $\delta$ are determined by Charlie's choice of \emph{source}. For instance, there might be sources $K$ that produce homogeneous ensembles, in which exactly one of the coefficients is 1 -- say, there is some mechanism that tends to align all the hidden spins in an ensemble.\footnote{In this case, a mixing protocol will have to be defined in such a way that pairs are selected and sent on to Alice and Bob one by one (to prevent the resulting mixture from homogenising), but arguably such a mechanism will work uniformly over all of Alice's and Bob's measurement contexts, so that mixtures will be operationally well-defined.} Note that if such sources exist and are reliable, then Charlie will be able to prepare ensembles that violate equiprobability to a maximal extent, since for instance an ensemble with $\alpha=1$ will yield $|\!\mean{A}_K^{IJ}\!|\approx1$ to an arbitary degree of approximation. Since it also reproduces the correlations required by Theorems~1 and 1$'$, it very clearly leads to signalling.

The Schulman model of course is not a complete theory. It contains no detailed treatment of how preparations or measurements work, and it does not even cover all quantum states. One might reasonably expect that distributions of hidden spins other than (\ref{Sch9}) or over different pairs of hidden spins might describe other quantum states. While this may be true, and models of this kind may provide us with candidates for $\psi$-epistemic theories\footnote{This is in fact the main line of argument in Wharton (2014).}, it is unimportant for our purposes, because distributions satisfying (\ref{Sch8}) but violating (\ref{Sch9}) do not reproduce the Born probabilities of \emph{any} quantum state. That is, even if in this model (or some appropriate extension) quantum states are not ontic but epistemic, the model includes non-quantum distributions that allow signalling in principle.\footnote{Ken Wharton, Rod Sutherland and co-workers are currently developing a novel hidden variables model violating MI, based on local weak values, which successfully describes arbitrary quantum circuits (Wharton \emph{et al.}\ 2024). It is not known yet whether quantum states are epistemic in this model.}

%%%%%%%%%%%%%%%%%%%%%%%
%                                                               %
%%%%%%%%%%%%%%%%%%%%%%%
\section{Experimental Metaphysics}\label{metaphysics}
Theorems~2 and 2$'$ assume violation of the Bell inequalities (including closing the conspiracy loophole), which may be considered amply established, as well as bounds on the correlations along certain chains of spin directions, for which a specific experiment has already been performed (with photons) by Stuart \emph{et al.}\ (2012). The experimental warrant for these assumptions is thus hardly in doubt, although it would be desirable to perform a combined loophole-free experiment in which both the test of the Bell inequalities and the measurement of the correlations along an appropriate chain of spin directions are carried out on the same ensemble. From Theorems~2 and 2$'$ it then follows (and robustly so) that any hidden variables models satisfying OI allow for signalling, \emph{provided it is possible in principle to prepare certain non-quantum ensembles}. The conclusion applies both to models in which PI is violated and to those in which MI is violated (or a combination of both), making all of these models testable in principle. 

Before we discuss in more detail the implications of our results, we should however ask whether the possibility of preparing such non-quantum ensembles is something to be generally expected: there may be hidden variables models in which Charlie cannot prepare the required ensembles even in principle. This might be the case also for hidden variables models that violate PI and satisfy MI (or violate both PI and MI),\footnote{For instance, as mentioned above, some authors dispute that disequilibrium distributions make sense in de~Broglie--Bohm theory (e.g.\ D\"{u}rr, Goldstein and Zangh\`{i} 1992). One might also doubt that they make sense in Nelson's stochastic mechanics, since the Born distribution is the unique joint solution of the forwards and backwards Fokker--Planck equations of the theory; but see Bacciagaluppi (2012) for how the notion of disequilibrium is applicable also in Nelson's theory.} but we are interested especially in those that violate MI and satisfy PI. Perhaps surprisingly, there are some rather well-known approaches to quantum mechanics that at least formally exhibit genuine violations of MI -- and in fact realise our toy model (\ref{12}) from the previous section -- but without allowing preparation of any ensembles other than quantum ones even in principle.

To set the stage, think of Bohr's (1935) reply to EPR. In Bohr's treatment, the EPR state is prepared by letting each of two particles with known total momentum pass through one of two slits at given distance from each other in a screen also with known momentum. Re-measuring the momentum of the screen yields a state for the particles with a known total momentum and the given difference in position. If Bob measures the momentum of his particle, he applies a physical picture in which the conservation laws hold and his particle has acquired a definite momentum by exchanging momentum with the screen (and indirectly with the other particle) at the time of preparation. He then observes this value at the time of measurement. Alternatively, if he measures the position of his particle, he applies a picture of spatio-temporal coordination in which his particle has acquired (as has Alice's particle) a definite position by passing through one of the slits at the time of preparation. He then observes this value at the time of measurement. Because they determine the results of the measurements, these values of momentum and position formally play the role of hidden variables. In fact, according to Hermann (1935), they show that quantum mechanics cannot be \emph{completed} via hidden variables because it already \emph{contains} the causes determining the measurement results. This description is irreducibly contextual, and the statistical distribution over the hidden variables is concentrated on different variables (momentum or position), depending on the context of measurement chosen. In this sense, while Bohr's analysis of quantum mechanical measurements is arguably not retrocausal, yet it systematically violates MI. 

In order to relate this to the Bell inequalities, note that Bohr himself in his reply to EPR adopts a single-user perspective (Bob either measures momentum or position, and possibly only later decides to measure the other quantity on Alice's particle), but one can consider a two-user perspective in which, say, Alice measures position and Bob measures momentum or -- to revert to our previous setting -- Alice measures spin along $A$ and Bob measures spin along $B$, and the \emph{total} context of measurement $IJ$ determines the distribution of the hidden variables, precisely as given by (\ref{12}). But Charlie has no more control over it than through the preparation of the initial quantum state. Thus, there is no signalling in principle even though the explanation of the distant correlations in a Bell setup relies on violation of MI.\footnote{This analysis of the bipartite case may not be strictly Bohrian, but it applies explicitly in approaches such as a neo-Bohrian reading of consistent histories as presented by Griffiths (2024, Sect.~6), who refers to it as `Copenhagen done right': given the total context of measurement $IJ$, one can consistently assign to the particle pair the projections $\ket{A=\pm1}\bra{A=\pm1}\otimes\ket{B=\pm1}\bra{B=\pm1}$ with statistical distribution (\ref{12}) even before Alice's and Bob's measurements.}

Another example of violation of MI without ability to prepare non-quantum distributions is provided by the Everett theory if one adopts the picture of \emph{divergent} worlds.\footnote{That is, the picture in which distinct but indistinguishable worlds exist even before they branch -- as opposed to the picture of \emph{splitting}, in which upon branching a single world splits into two or more worlds. (There are arguably slightly different versions of this distinction, but here we follow Wilson (2020). Thanks to Paul Tappenden for helpful correspondence on this point.)} One often says that Bell's theorem does not apply to Everett, but of course \emph{in individual Everett worlds} one predicts that in Bell experiments the Bell inequalities will be violated, and those predictions are confirmed (except in some deviant worlds), so that one of OI, PI or MI must have been violated. If one takes it that in Everett there are no variables in addition to the quantum state (of a world), then MI cannot be violated, nor (again except in some deviant worlds) can PI. Thus, Bell inequality violations in Everett will be naturally interpreted in terms of violations of OI. But if one adopts the picture of divergence, then even before Alice's and Bob's measurements take place there are different sets of worlds, each characterised by the way they later evolve into worlds with the outcomes $(A=\pm1,B=\pm1)$, respectively, and which diverge from the others for instance along the future light cones of Alice's and Bob's measurements.\footnote{For a detailed discussion of relativistic aspects of branching (and an argument to the effect that the distinction between divergence and splitting is not metaphysically substantive), see Bacciagaluppi (2026).} These spin values are formally hidden variables that determine the outcomes in the respective worlds, and in every non-deviant world  their statistical distribution over an ensemble of particle pairs is given by (\ref{12}). But Charlie has no control over this distribution beyond that provided by the preparation of the quantum state (again perhaps except in certain deviant worlds). We thus have no signalling in principle. 

While at first striking, the fact that even though these approaches violate MI they do not allow for signalling in principle need not really be surprising, because the dependence of the hidden variables on Alice's and Bob's settings is arguably not causal. Even in Bohr's case (to echo his own phrases), the settings do not determine the distribution of hidden variables in any `mechanical' sense, rather they determine the conditions under which different objective physical descriptions are applicable. It may also be significant that the introduction of hidden variables in these examples is merely formal: by moving the Heisenberg cut or by adopting the picture of splitting, respectively, the complete state of the particle pair becomes again the quantum state, so that MI is trivially satisfied. We do not pretend to have any general arguments to this effect and do not wish to enter debates about causation, but it is possible that in all cases in which the hidden variables can be said to \emph{causally} depend on Alice's and Bob's settings, Charlie has in principle the ability to prepare the required non-quantum distributions, thus enabling at least one of Alice or Bob to signal.\footnote{Note that in their review of superdeterministic theories, Hossenfelder and Palmer (2020, Sect.~6) also argue that, at least as regards short-term trends, one might expect in principle to be able to prepare ensembles that violate the Born rule (thus in our setting arguably to signal).}

These conclusions appear at first to be contradicted by the arguments of Sen and Valentini (2020a,b), who claim that no genuine signalling is possible in `superdeterminist' theories -- by which they mean theories that are deterministic and that fix both the hidden variables and the settings either via common causes or global constraints (while our theorems hold irrespective of whether one imagines the underlying theory to be superdeterminist in this sense). The disagreement, however, is explained by the fact that we define signalling in operational terms, while Sen and Valentini make a finer distinction between genuine signalling and `apparent signalling', on the basis of a causal analysis at a more fundamental level, and using a specific notion of causation.\footnote{Sen and Valentini take causation in a deterministic theory to be expressed by functional dependence. They rightly point out that in pilot-wave theory there is direct functional dependence of the outcomes on the settings via action at a distance (as we sketched in Section~\ref{signalling}), while in superdeterminist theories in their sense the dependence arises solely at the level of statistical distributions and contingently on the initial conditions (not only in the case of common causes but crucially also in that of global constraints). In this paper, we do not commit to any specific notion of causation, but we are unconvinced that functional dependence is appropriate in this setting: its relation to causation is notoriously ambiguous (as pointed out long ago by Russell!), and insofar as it is a useful tool in assessing causal dependence, it seems to us that it is applicable only after the identification of free variables and thus possibly only at an emergent level and contingently on initial conditions.} 

With these qualifications in place, our results can be taken as an extension of  Shimony's programme of `experimental metaphysics'. Shimony (1989) proposed that the violations of the Bell inequalities allow us to draw conclusions from experimental results to the nature of nonlocality, irrespective of the validity of quantum theory. Originally, this claim was limited to nonlocality in the form of violations of OI or PI. Accordingly, nonlocality would be compatible with relativity or not depending on whether signalling in principle is ruled out (violations of OI) or not (violations of PI).\footnote{Some authors dispute this final point and argue that even violations of OI are incompatible with (the causal constraints implied by) relativity (Maudlin 1994, Norsen 2009). We side with Shimony, and regard his view as mainstream, but will not argue for it here (for that, see Myrvold (2016) and Bacciagaluppi (in preparation).} Once standard caveats are taken into account,\footnote{For one, Jones and Clifton (1993) point out that in some cases where both OI and PI fail, the violation of the latter can be mediated by violation of the former. These are cases in which violations of PI are compatible with relativity. Therefore, the claims of experimental metaphysics should be read at most as applying to cases in which only OI is violated or only PI is violated, but not both. We have also made no particular claims about cases in which also OI is violated. Further, making conclusive claims about relativity and about signalling is generally only possible if the model under consideration is taken to be fundamental. As mentioned, it is clear that a model violating OI (e.g.\ a model based on standard quantum mechanics) could be explained through one violating PI (e.g.\ a model based on pilot-wave theory); thus, compatibility with relativity at one level might be lost at a deeper level. And we have also pointed out already that violations of PI may correspond to violations of MI in a more detailed model; thus, reversing the point just made, a model that appears to be incompatible with relativity at one level may after all be compatible with relativity at a deeper level.}
and now that the experimental loopholes are convincingly filled, Shimony's argument is sound, \emph{provided one assumes MI}.

Our results cover this final proviso by establishing precise conditions under which violations of MI lead to signalling and thus \emph{also} lead to a form of nonlocality. In pilot-wave theory, the notion of disequilibrium ensembles opens up the possibility of new physics, with a number of experiments having been proposed to detect it.\footnote{For a recent overview, see Valentini (2024).} So does the notion of non-quantum ensembles for theories violating MI. Importantly, signalling due to action at a distance and signalling due to violations of MI can in principle be distinguished operationally.\footnote{Thereby answering a criticism raised by Norsen in dialogue with Price (Norsen and Price 2021, Sect.~5).} Action at a distance arguably requires a preferred time ordering between the measurement settings and the distant outcomes: for Alice or Bob to be able to signal to the other, it is necessary that they perform their measurement \emph{first}. There is no such restriction on signals arising through violations of MI, which require no preferred time ordering. Signalling in principle is thus the crucial notion that makes it possible to distinguish operationally between the different forms of nonlocality: correlations without signalling, signalling with a preferred time order, or signalling with no preferred time order. The new form of nonlocality associated with violations of MI is stronger than the nonlocality associated with violations of OI, because it allows signalling in principle, but it is milder than the action at a distance traditionally associated with violations of PI, because it is arguably compatible with relativity.

%%%%%%%%%%%%%%%%%%%%%%%
%                                                               %
%%%%%%%%%%%%%%%%%%%%%%%
\begin{appendix}
\section{Proofs of Theorems~0 and 0$'$}\label{A}
Take a hidden variables model of bipartite spin measurements, and consider two finite chains of coplanar spin-$\frac{1}{2}$ observables $A_i$,~$B_j$ ($i=0,\ldots n$; $j=0,\ldots,n-1$), such that the angle between $A_i$ and $A_{i+i}$ is $\frac{\theta}{n}$ and $B_j$ is halfway between $A_j$ and $A_{j+1}$ (see again Figure~2).\footnote{This proof is based on a step in Leegwater's proof of the Colbeck--Renner theorem (Leegwater 2016, Sect.~4).} Assuming PI, we have
\begin{subequations}
  \begin{align}
  p_\lambda^{I_i}(A_i=+1):\!\!&= p_\lambda^{I_iJ_j}(A_i=+1,B_j=+1)+p_\lambda^{I_iJ_j}(A_i=+1,B_j=-1)\;=\label{app1a}\\
                              &= p_\lambda^{I_iJ_{j'}}(A_i=+1,B_{j'}=+1)+p_\lambda^{I_iJ_{j'}}(A_i=+1,B_{j'}=-1)
  \end{align}\label{app1}
\end{subequations}
and
\begin{subequations}
  \begin{align}
  p_\lambda^{J_j}(B_j=+1):\!\!&= p_\lambda^{I_iJ_j}(A_i=+1,B_j=+1)+p_\lambda^{I_iJ_j}(A_i=-1,B_j=+1)\;=\label{app2a}\\
                              &= p_\lambda^{I_{i'}J_j}(A_{i'}=+1,B_j=+1)+p_\lambda^{I_{i'}J_j}(A_{i'}=-1,B_j=+1) \ .
  \end{align}\label{app2}
\end{subequations}
And of course (e.g.\ subtracting (\ref{app2a}) from (\ref{app1a}))
\begin{multline}
  |p_\lambda^{I_i}(A_i=+1)-p_\lambda^{J_j}(B_j=+1)|\,=\,|p_\lambda^{I_iJ_j}(A_i=-1,B_j=+1)-p_\lambda^{I_iJ_j}(A_i=+1,B_j=-1)|\;\;\leq\\
                                                             \leq\;\;|p_\lambda^{I_iJ_j}(A_i=-1,B_j=+1)|+|p_\lambda^{I_iJ_j}(A_i=+1,B_j=-1)|\,=\, p_\lambda^{I_iJ_j}(A_i\neq B_j) \ .
  \label{app3}
\end{multline}
From (\ref{app1})--(\ref{app3}) and the triangle inequality we obtain
  \begin{align}
    |p_\lambda^{I_0}(A_0=+1)-p_\lambda^{I_n}(A_n=+1)| & \leq \nonumber\\ 
   & \leq \sum_{i=0}^{n-1}|p_\lambda^{I_i}(A_i=+1)-p_\lambda^{J_i}(B_i=+1)| + 
          \sum_{i=0}^{n-1}|p_\lambda^{J_i}(B_i=+1)-p_\lambda^{I_{i+1}}(A_{i+1}=+1)| \leq  \nonumber \\
                                                                                    &
    \leq \sum_{i=0}^{n-1}p_\lambda^{I_iJ_i}(A_i\neq B_i) + \sum_{i=0}^{n-1}p_\lambda^{I_{i+1}J_i}(B_i\neq A_{i+1}) \ . \label{app3b}
%\nonumber 
  \end{align}
That is,
  \begin{multline}
-\sum_{i=0}^{n-1}p_\lambda^{I_iJ_i}(A_i\neq B_i) - \sum_{i=0}^{n-1}p_\lambda^{I_{i+1}J_i}(B_i\neq A_{i+1})
\;\leq \;p_\lambda^{I_0}(A_0=+1)-p_\lambda^{I_n}(A_n=+1)
\;\leq \; \\ 
\leq\;\sum_{i=0}^{n-1}p_\lambda^{I_iJ_i}(A_i\neq B_i) + \sum_{i=0}^{n-1}p_\lambda^{I_{i+1}J_i}(B_i\neq A_{i+1}) \ . \label{app3bb}
  \end{multline}
By substituting $-B_i$ for $B_i$ in  (\ref{app3bb}) we have similarly
  \begin{multline}
-\sum_{i=0}^{n-1}p_\lambda^{I_iJ_i}(A_i= B_i) - \sum_{i=0}^{n-1}p_\lambda^{I_{i+1}J_i}(B_i= A_{i+1})
\;\leq \;p_\lambda^{I_0}(A_0=+1)-p_\lambda^{I_n}(A_n=+1)
\;\leq \; \\ 
\leq\;\sum_{i=0}^{n-1}p_\lambda^{I_iJ_i}(A_i= B_i) + \sum_{i=0}^{n-1}p_\lambda^{I_{i+1}J_i}(B_i= A_{i+1}) \ . \label{app3c}
  \end{multline}
Now consider an ensemble $K$, and, assuming MI, average in (\ref{app3bb}) and (\ref{app3c}) with $\sigma_K(\lambda)$. Note that, for any 
 $i,j$, (\ref{corr}) is equivalent to
  \begin{equation}
 %   \int\sigma_K^{I_iJ_j}(\lambda)d\lambda\,p_\lambda^{I_iJ_j}(A_i\neq B_j)=
    \int\sigma_K(\lambda)d\lambda\,p_\lambda^{I_iJ_j}(A_i\neq B_j)
    \leq\delta \ , \label{app38}
  \end{equation}
and (\ref{anticorr}) is equivalent to  
  \begin{equation}
%    \int\sigma_K^{I_iJ_j}(\lambda)d\lambda\,p_\lambda^{I_iJ_j}(A_i=B_j)=
    \int\sigma_K(\lambda)d\lambda\,p_\lambda^{I_iJ_j}(A_i=B_j)
    \leq\delta \ .
  \end{equation}
If on average either (\ref{corr}) holds for all links in the chain or (\ref{anticorr}) holds for all links in the chain, it follows from (\ref{app3bb}) and (\ref{app3c}) that 
  \begin{equation} 
    \Big|\int\sigma_K(\lambda)d\lambda\,\Big(p_\lambda^{I_0}(A_0=+1)-p_\lambda^{I_n}(A_n=+1)\Big)\Big|
         \;\leq\; 2n\delta \ .
  \end{equation}
In the case $\theta=\pi$, we have $A_n=+1\iff A_0=-1$, and thus
  \begin{equation} 
    |\mean{A_0}_K^{I_0}| \leq 2n\delta \ . \label{app41}
  \end{equation}
 This proves Theorem 0$'$.

For the proof of Theorem 0, note that if $K$ reproduces the (anti)\-correlations of a maximally entangled state, then we can set $\delta=\tfrac{\theta^2}{16n^2}$ (in fact $\delta=\sin^2\Big(\tfrac{\theta}{4n}\Big)$) in (\ref{app38})--(\ref{app41}). Taking the limit $n\rightarrow\infty$, this yields
  \begin{equation}
    p_\lambda^{I_0}(A_0=+1)\;=\;p_\lambda^{I_n}(A_n=+1) \ ,
  \end{equation}
up to a measure-zero set of $\lambda$s and independently of $I_0$ and $I_n$. In particular, for $\theta=\pi$, it yields $|\mean{A_0}_K^{I_0}|=0$.

\section{Direct Proofs of Theorems 1 and 1$'$}\label{B}
As above, take a hidden variables model of bipartite spin measurements, and consider two finite chains of coplanar spin-$\frac{1}{2}$ observables $A_i$,~$B_j$ ($i=0,\ldots n$; $j=0,\ldots,n-1$), such that the angle between $A_i$ and $A_{i+i}$ is $\frac{\theta}{n}$ and $B_j$ is halfway between $A_j$ and $A_{j+1}$. Consider an ensemble $K$ and the corresponding probabilities (\ref{Kprobs}). Assuming that neither Alice nor Bob can signal, we have
\begin{subequations}
  \begin{align}
  p_K^{I_i}(A_i=+1):\!\!&= p_K^{I_iJ_j}(A_i=+1,B_j=+1)+p_K^{I_iJ_j}(A_i=+1,B_j=-1)\;=\label{app1aK}\\
                              &= p_K^{I_iJ_{j'}}(A_i=+1,B_{j'}=+1)+p_K^{I_iJ_{j'}}(A_i=+1,B_{j'}=-1)
  \end{align}\label{app1K}
\end{subequations}
and
\begin{subequations}
  \begin{align}
  p_K^{J_j}(B_j=+1):\!\!&= p_K^{I_iJ_j}(A_i=+1,B_j=+1)+p_K^{I_iJ_j}(A_i=-1,B_j=+1)\;=\label{app2aK}\\
                              &= p_K^{I_{i'}J_j}(A_{i'}=+1,B_j=+1)+p_K^{I_{i'}J_j}(A_{i'}=-1,B_j=+1) \ .
  \end{align}\label{app2K}
\end{subequations}
And, subtracting (\ref{app2aK}) from (\ref{app1aK}),
\begin{multline}
  |p_K^{I_i}(A_i=+1)-p_K^{J_j}(B_j=+1)|\,=\,|p_K^{I_iJ_j}(A_i=-1,B_j=+1)-p_K^{I_iJ_j}(A_i=+1,B_j=-1)|\;\;\leq\\
                                                             \leq\;\;|p_K^{I_iJ_j}(A_i=-1,B_j=+1)|+|p_K^{I_iJ_j}(A_i=+1,B_j=-1)|\,=\, p_K^{I_iJ_j}(A_i\neq B_j) \ .
  \label{app3K}
\end{multline}
From (\ref{app1K})--(\ref{app3K}) and the triangle inequality we obtain
  \begin{align}
    |p_K^{I_0}(A_0=+1)-p_K^{I_n}(A_n=+1)| & \leq \nonumber\\ 
   & \leq \sum_{i=0}^{n-1}|p_K^{I_i}(A_i=+1)-p_K^{J_i}(B_i=+1)| + 
          \sum_{i=0}^{n-1}|p_K^{J_i}(B_i=+1)-p_K^{I_{i+1}}(A_{i+1}=+1)| \leq  \nonumber \\
                                                                                    &
    \leq \sum_{i=0}^{n-1}p_K^{I_iJ_i}(A_i\neq B_i) + \sum_{i=0}^{n-1}p_K^{I_{i+1}J_i}(B_i\neq A_{i+1}) \ . \label{app3bK}
%\nonumber 
  \end{align}
That is,
  \begin{multline}
-\sum_{i=0}^{n-1}p_K^{I_iJ_i}(A_i\neq B_i) - \sum_{i=0}^{n-1}p_K^{I_{i+1}J_i}(B_i\neq A_{i+1})
\;\leq \;p_K^{I_0}(A_0=+1)-p_K^{I_n}(A_n=+1)
\;\leq \; \\ 
\leq\;\sum_{i=0}^{n-1}p_K^{I_iJ_i}(A_i\neq B_i) + \sum_{i=0}^{n-1}p_K^{I_{i+1}J_i}(B_i\neq A_{i+1}) \ . \label{app3bbK}
  \end{multline}
By substituting $-B_i$ for $B_i$ in (\ref{app3bbK}) we have, of course,
  \begin{multline}
-\sum_{i=0}^{n-1}p_K^{I_iJ_i}(A_i= B_i) - \sum_{i=0}^{n-1}p_K^{I_{i+1}J_i}(B_i= A_{i+1})
\;\leq \;p_K^{I_0}(A_0=+1)-p_K^{I_n}(A_n=+1)
\;\leq \; \\ 
\leq\;\sum_{i=0}^{n-1}p_K^{I_iJ_i}(A_i= B_i) + \sum_{i=0}^{n-1}p_K^{I_{i+1}J_i}(B_i= A_{i+1}) \ . \label{app3cK}
  \end{multline}
If on average $K$ satisfies (\ref{corr}) (i.e.\ $p_K^{I_iJ_j}(A_i\neq B_j)\leq\delta$) for all links in the chain, or (\ref{anticorr}) (i.e.\ $p_K^{I_iJ_j}(A_i= B_j)\leq\delta$) for all links in the chain, it follows from (\ref{app3bbK}) and (\ref{app3cK}) that 
  \begin{equation} 
    |p_K^{I_0}(A_0=+1)-p_K^{I_n}(A_n=+1)|
         \;\leq\; 2n\delta \ .
  \end{equation}
In the case $\theta=\pi$, we have $A_n=+1\iff A_0=-1$, and thus
  \begin{equation} 
    |\mean{A_0}_K^{I_0}| \leq 2n\delta \ , \label{app41K}
  \end{equation}
which completes the proof of Theorem~1$'$.

For Theorem 1, if $K$ reproduces the (anti)\-correlations of a maximally entangled state, we can insert $\delta=\tfrac{\theta^2}{16n^2}$; and taking the limit $n\rightarrow\infty$, we obtain
  \begin{equation}
    p_K^{I_0}(A_0=+1)\;=\;p_K^{I_n}(A_n=+1) \ ,
  \end{equation}
independently of $I_0$ and $I_n$, and in particular $|\mean{A_0}_K^{I_0}|=0$.

\section{Proofs of Lemmas 1 and 2}\label{C}
Let Charlie prepare an ensemble $K$ of pairs of spin-$\frac{1}{2}$ particles, with Alice 
and Bob measuring spins $A,A',B,B'$ in the contexts $I,I',J,J'$, respectively. Assume a 
hidden variables model of this setup that satisfies OI. This means that
  \begin{equation}
    \langle AB\rangle^{IJ}_\lambda + \langle A'B\rangle^{I'J}_\lambda + 
    \langle AB'\rangle^{IJ'}_\lambda - \langle A'B'\rangle^{I'J'}_\lambda  
  \end{equation}
becomes 
  \begin{equation}
     \langle A\rangle^{IJ}_\lambda\langle B\rangle^{IJ}_\lambda + 
     \langle A'\rangle^{I'J}_\lambda\langle B\rangle^{I'J}_\lambda +
     \langle A\rangle^{IJ'}_\lambda\langle B'\rangle^{IJ'}_\lambda -
     \langle A'\rangle^{I'J'}_\lambda\langle B'\rangle^{I'J'}_\lambda 
  \end{equation}
(and similarly with the minus sign positioned differently). 

For any $I,J$ let $\sigma_1^{IJ}(\lambda)$ be the sub-distribution of $\sigma_K^{IJ}(\lambda)$ consisting of all pairs with hidden variables $\lambda$ such that $\langle A\rangle^{IJ}_\lambda\geq 0$. Similarly, let $\sigma_2^{IJ}(\lambda)$ be the sub-distribution such that $\langle A\rangle^{IJ}_\lambda< 0$.

We have:
\begin{align}  
  \Big|\!\int\!\!\sigma_K^{IJ}(\lambda)d\lambda\langle A\rangle^{IJ}_\lambda\langle B\rangle^{IJ}_\lambda\Big|\; &= \;\Big|\alpha\!\!\int\!\!\sigma_1^{IJ}(\lambda)d\lambda\langle A\rangle^{IJ}_\lambda\langle B\rangle^{IJ}_\lambda\;+\;
               (1\!-\!\alpha)\!\!\int\!\!\sigma_2^{IJ}(\lambda)d\lambda\langle A\rangle^{IJ}_\lambda\langle B\rangle^{IJ}_\lambda\Big|\;\leq
                                                                                                                                               \nonumber \\
         &\leq\;\alpha\!\!\int\!\!\sigma_1^{IJ}(\lambda)d\lambda\;\big|\langle A\rangle^{IJ}_\lambda\langle B\rangle^{IJ}_\lambda\big|\;+\;
               (1\!-\!\alpha)\!\!\int\!\!\sigma_2^{IJ}(\lambda)d\lambda\;\big|\langle A\rangle^{IJ}_\lambda\langle B\rangle^{IJ}_\lambda\big|\;\leq
                                                                                                                                               \nonumber \\
         &\leq\;\left(\alpha\!\!\int\!\!\sigma_1^{IJ}(\lambda)d\lambda\;\big|\langle A\rangle^{IJ}_\lambda\big|\;+\;
               (1\!-\!\alpha)\!\!\int\!\!\sigma_2^{IJ}(\lambda)d\lambda\;\big|\langle A\rangle^{IJ}_\lambda\big|\right)\cdot\max_\lambda\big|\langle B\rangle^{IJ}_\lambda\big|\;\leq
                                                                                                                                               \nonumber \\
         &\leq\;\alpha\!\!\int\!\!\sigma_1^{IJ}(\lambda)d\lambda\;\big|\langle A\rangle^{IJ}_\lambda\big|\;+\;
               (1\!-\!\alpha)\!\!\int\!\!\sigma_2^{IJ}(\lambda)d\lambda\;\big|\langle A\rangle^{IJ}_\lambda\big|\;=
                                                                                                                                               \nonumber  \\
         &=\;\alpha\!\!\int\!\!\sigma_1^{IJ}(\lambda)d\lambda\;\langle A\rangle^{IJ}_\lambda\;-\;
               (1\!-\!\alpha)\!\!\int\!\!\sigma_2^{IJ}(\lambda)d\lambda\;\langle A\rangle^{IJ}_\lambda\;=
                                                                                                                                               \nonumber  \\
         &=\;\alpha\langle A\rangle^{IJ}_{1}\;-\;(1\!-\!\alpha)\langle A\rangle^{IJ}_{2}\;=      \nonumber \\[1.0ex]
         &=  \;  \alpha|\langle A\rangle^{IJ}_{1}|\;+\;(1\!-\!\alpha)|\langle A\rangle^{IJ}_{2}| \ .
         \end{align}

Now assume that for all subdistributions $\sigma_0^{IJ}(\lambda)$ with weight 
     $\frac{1}{2}\leq\alpha\leq 1$ in $\sigma_{K}^{IJ}(\lambda)$,
       \begin{equation}
         \big|\mean{A}^{IJ}_0\big|\leq\varepsilon  
         \label{new1}
       \end{equation}   
for some $0\leq\varepsilon<1$. Note that, since $\sigma_1^{IJ}(\lambda)$ and $\sigma_2^{IJ}(\lambda)$ exhaust $\sigma_{K}^{IJ}$, at least one of them has weight at least $\frac{1}{2}$. We assume it is $\sigma_1^{IJ}(\lambda)$, but this is immaterial for the proof.

It follows that
  \begin{equation}
    \alpha|\langle A\rangle^{IJ}_{1}|\;+\;(1\!-\!\alpha)|\langle A\rangle^{IJ}_{2}|\leq \alpha\varepsilon+(1-\alpha)=
    1-\alpha(1-\varepsilon)\leq 1-\tfrac{1}{2}(1-\varepsilon)=
    \tfrac{1}{2}\varepsilon+\tfrac{1}{2}\ .
  \end{equation}

Therefore, if we assume (\ref{new1}) for both $A$ in the contexts $IJ$ and $IJ'$, and $A'$ in the contexts $I'J$ and $I'J'$ (or similarly for both $B$ in the contexts $IJ$ and $I'J$, and $B'$ in the contexts $IJ'$ and $I'J'$), then
\begin{align}
  \Big|\!\int\!\!\sigma_K^{IJ}(\lambda)d\lambda\,\langle A\rangle^{IJ}_\lambda&\langle B\rangle^{IJ}_\lambda\;+\; 
  \!\int\!\!\sigma_K^{I'J}(\lambda)d\lambda\,\langle A'\rangle^{I'J}_\lambda\langle B\rangle^{I'J}_\lambda\;+\nonumber \\
 &+ \;\int\!\!\sigma_K^{IJ'}(\lambda)d\lambda\,\langle A\rangle^{IJ'}_\lambda\langle B'\rangle^{IJ'}_\lambda\;-\;
  \int\!\!\sigma_K^{I'J'}(\lambda)d\lambda\,\langle A'\rangle^{I'J'}_\lambda\langle B'\rangle^{I'J'}_\lambda\Big|\leq\nonumber\\[1.5ex]
  \Big|\!\int\!\!\sigma_K^{IJ}(\lambda)d\lambda\,\langle A\rangle^{IJ}_\lambda&\langle B\rangle^{IJ}_\lambda\Big|\;+\; 
  \Big|\int\!\!\sigma_K^{I'J}(\lambda)d\lambda\,\langle A'\rangle^{I'J}_\lambda\langle B\rangle^{I'J}_\lambda\Big|\;+\nonumber \\
 &+ \Big|\int\!\!\sigma_K^{IJ'}(\lambda)d\lambda\,\langle A\rangle^{IJ'}_\lambda\langle B'\rangle^{IJ'}_\lambda\Big|\;+\;
  \Big|\int\!\!\sigma_K^{I'J'}(\lambda)d\lambda\,\langle A'\rangle^{I'J'}_\lambda\langle B'\rangle^{I'J'}_\lambda\Big|\;\leq\;2\varepsilon + 2
  \label{app43}
\end{align}
(and similary with the minus sign positioned differently). Lemma~1 follows by contraposition.

For the proof of Lemma~2, take any $A,B$ with corresponding contexts $I,J$, and consider 
  \begin{equation}
    \int\sigma_K^{IJ}(\lambda)d\lambda\, p^{IJ}_\lambda(A\neq B) \ .
  \end{equation} 
Take any sub-distributions $\sigma_1^{IJ}(\lambda)$ and $\sigma_2^{IJ}(\lambda)$ such that
  \begin{equation}
    \sigma_K^{IJ}(\lambda)=\alpha\sigma_1^{IJ}(\lambda)
    +(1-\alpha)\sigma_2^{IJ}(\lambda) \ ,
  \end{equation}
with $0<\alpha\leq1$. Then, by definition,
   \begin{equation}
     \alpha\int\sigma_1^{IJ}(\lambda)d\lambda\, p^{IJ}_\lambda(A\neq B)
     +
     (1-\alpha)\int\sigma_2^{IJ}(\lambda)d\lambda 
     \, p^{IJ}_\lambda(A\neq B)=\int\sigma_K^{IJ}(\lambda)d\lambda\, p^{IJ}_\lambda(A\neq B) \ ,
   \end{equation}
and therefore 
  \begin{multline}
    \int\sigma_1^{IJ}(\lambda)d\lambda
     \, p^{IJ}_\lambda(A\neq B)=\frac{1}{\alpha}\Big(
     \int\sigma_K^{IJ}(\lambda)d\lambda\, p^{IJ}_\lambda(A\neq B)
     -(1-\alpha)\int\sigma_2^{IJ}(\lambda)d\lambda 
     \, p^{IJ}_\lambda(A\neq B)\Big)\leq \\
     \leq \frac{1}{\alpha}\int\sigma_K^{IJ}(\lambda)d\lambda\, p^{IJ}_\lambda(A\neq B)
     \ .
  \end{multline}
For positively correlated $A$ and $B$, the claim follows because $\mean{AB}_\lambda^{IJ}=1-2p_\lambda^{IJ}(A\neq B)$. For negatively correlated $A$ and $B$, repeat the proof substituting $-B$ for $B$.

\end{appendix}

%%%%%%%%%%%%%%%%%%%%%%%
%                                                               %
%%%%%%%%%%%%%%%%%%%%%%%
\section*{\normalsize Acknowledgements}
{\small
Previous versions of this work were presented in 2018 at \emph{Quantum Contextuality in Quantum Mechanics and Beyond} in Prague and at \emph{Foundations} in Utrecht, and in 2023 at Merton College, Oxford, at \emph{Foundations} in Bristol and for the Harvard Foundations Seminar. The final version was presented at the EPSA satellite meeting on retrocausation in Groningen in August 2025. The authors would like to thank the various audiences at these talks, as well as Harvey Brown, Roger Colbeck, Huw Price, Paul Tappenden, Ken Wharton and several others for discussions or correspondence, especially Wayne Myrvold for some extremely useful comments and for pointing out a mistake in a previous version of Lemma 1. RH's research during the initial stages of the project was supported by NWO Veni Grant No.~275-20-070, while part of the redaction of this paper was carried out while GB was Benjamin Meaker Distinguished Visiting Professor in the Department of Philosophy at the University of Bristol. We gratefully acknowledge this support and hospitality.
}

%%%%%%%%%%%%%%%%%%%%%%%
%                                                               %
%%%%%%%%%%%%%%%%%%%%%%%
%\nocite{*} % prints the entire bibliography
%\printbibliography
%
%\
%
%\
% bib file is defined in packages_and_options.tex

\section*{\normalsize References}
{\small
\noindent Almada, D., Ch'ng, K., Kintner, S., Morrison, B., and Wharton, K. (2016), `Are retrocausal accounts of entanglement unnaturally fine-tuned?', \emph{International Journal of Quantum Foundations} {\bf 2}, 1--16. \url{https://arxiv.org/pdf/1510.03706}

\

\noindent Bacciagaluppi, G. (2012), `Non-equilibrium in stochastic mechanics', \emph{Journal of Physics: Conference Series} {\bf 361}, 012017/1--12. \url{http://philsci-archive.pitt.edu/9120/}

\

\noindent Bacciagaluppi, G. (2026), `The relativity of branching', in A.~Ney (ed.), \emph{Local Quantum Mechanics: Everett, Many Worlds, and Reality} (Oxford: OUP), in press. \url{https://philsci-archive.pitt.edu/25071/}

\

\noindent Bacciagaluppi, G. (in preparation), `Against ``local causality'''. \url{https://www.dpg-verhandlungen.de/year/2025/conference/bonn/static/syqt1.pdf}

\

\noindent Barrett, J[eff] (1999), \emph{The Quantum Mechanics of Minds and Worlds} (Oxford: OUP). \url{https://sites.socsci.uci.edu/~jabarret/bio/publications/bookpdf.pdf}

\

\noindent Barrett, J[on], and Gisin, N. (2011), `How much measurement independence is needed to demonstrate
non\-locality?', \emph{Physical Review Letters} {\bf 106}, 100406/1--4. \url{https://arxiv.org/pdf/1008.3612}

\

\noindent Barrett, J[on], Kent, A., and Pironio, S. (2006), `Maximally nonlocal and monogamous quantum correlations', \emph{Physical Review Letters} {\bf 97}, 170409/1--4. \url{https://arxiv.org/pdf/quant-ph/0605182}

\

\noindent Bell, J. S. (1981), `Bertlmann's socks and the nature of reality', \emph{Journal de Physique} {\bf 42}, C2/41--61. Repr.\ as Chap.~16 in \emph{Speakable and Unspeakable in Quantum Mechanics} (Cambridge: CUP, 1987), pp.~139--158. \url{https://hal.science/jpa-00220688/document}

\

\noindent Bohr, N. (1935), `Can quantum-mechanical description of physical reality be considered complete?', \emph{Physical Review} {\bf 48}(8), 696--702. \url{https://journals.aps.org/pr/pdf/10.1103/PhysRev.48.696}

\

\noindent Butterfield, J. (1992), `Bell's theorem: What it takes', \emph{The British Journal for the Philosophy of Science} {\bf 43}(1), 41--83. \url{https://www.researchgate.net/publication/245583158_Bell's_Theorem_What_it_Takes}

\

\noindent Colbeck, R., and Renner, R. (2011), `No extension of quantum theory can have improved predictive power', \emph{Nature Communications} {\bf 2}(1), 411/1--5. \url{https://www.nature.com/articles/ncomms1416.pdf}

\

\noindent Colbeck, R., and Renner, R. (2016), `The completeness of quantum theory for predicting measurement outcomes', in G.~Chiribella and R.~Spekkens (eds), \emph{Quantum Theory: Informational Foundations and Foils} (Dordrecht: Springer), pp.~497--528. \url{https://arxiv.org/pdf/1208.4123}

\

\noindent D\"{u}rr, D., Goldstein, S., and Zangh\`{i}, N. (1992), `Quantum equilibrium and the origin of absolute uncertainty', \emph{Journal of Statistical Physics} {\bf 67}(5), 843--907. \url{https://arxiv.org/pdf/quant-ph/0308039}

\

\noindent Earman, J., and Norton, J. (1998), `Exorcist XIV: The wrath of Maxwell’s demon. Part I. From Maxwell to Szilard', \emph{Studies in History and Philosophy of Modern Physics} {\bf 29}(4), 435--471. \url{https://sites.pitt.edu/~jdnorton/papers/ExorcistXIV/Exorcist1.pdf}

\

\noindent Earman, J., and Norton, J. (1999), `Exorcist XIV: The wrath of Maxwell’s demon. Part II. From Szilard to Landauer and beyond', \emph{Studies in History and Philosophy of Modern Physics} {\bf 30}(1), 1--40. \url{https://sites.pitt.edu/~jdnorton/papers/ExorcistXIV/Exorcist2.pdf}

\

\noindent Friedman, A., Guth, A., Hall, M., Kaiser, D., and Gallicchio, J. (2019), `Relaxed Bell inequalities with arbitrary measurement dependence for each observer', \emph{Physical Review A} {\bf 99}, 012121/1--23. \url{https://arxiv.org/pdf/1809.01307}

\

\noindent Griffiths, R. (2024), `The consistent histories approach to quantum mechanics', in E.~Zalta and U.~Nodelman (eds), \emph{The Stanford Encyclopedia of Philosophy (Summer 2024 Edition)}. \url{https://plato.stanford.edu/archives/sum2024/entries/qm-consistent-histories/}

\

\noindent Hall, M. (2011), `Relaxed Bell inequalities and Kochen-Specker theorems', \emph{Physical Review A} {\bf 84}, 022102/1--16. \url{https://arxiv.org/pdf/1102.4467}

\

\noindent Hermann, G. (1935), `Die naturphilosophischen Grundlagen der Quantenmechanik', \emph{Abhandlungen der
Fries’schen Schule (Neue Folge)}, {\bf 6}(2), 69–152. Translated as `The natural-philosophical foundations of quantum mechanics',  in E.~Crull and G.~Bacciagaluppi (eds), \emph{Grete Hermann: Between Physics and Philosophy} (Dordrecht: Springer), pp.~239--278. 

\

\noindent Hermens, R. (2019), `An operationalist perspective on setting dependence', \emph{Foundations of Physics} {\bf 49}(3), 260--282. \url{https://link.springer.com/article/10.1007/s10701-019-00243-5}

\

\noindent Hermens, R. (2020), `Completely real? A critical note on the claims by Colbeck and Renner', \emph{Studies in History and Philosophy of Modern Physics} {\bf 72}, 121--137. \url{https://arxiv.org/pdf/2008.01444}

\

\noindent 't~Hooft, G. (2016), \emph{The Cellular Automaton Interpretation of Quantum Mechanics} (Cham: Springer). \url{https://link.springer.com/book/10.1007/978-3-319-41285-6}

\

\noindent Hossenfelder, S., and Palmer, T. (2020), `Rethinking superdeterminism', \emph{Frontiers in Physics} {\bf 8}, 139/1--13. \url{https://www.frontiersin.org/journals/physics/articles/10.3389/fphy.2020.00139/full}

\

\noindent Jones, M., and Clifton, R. (1993), `Against experimental metaphysics', \emph{Midwest Studies in Philosophy} {\bf 18}, 295--316. \url{https://onlinelibrary.wiley.com/doi/pdf/10.1111/j.1475-4975.1993.tb00269.x}

\

\noindent Kaiser, D. (2022), `Tackling loopholes in experimental tests of Bell's inequality', in O.~Freire, G.~Bacciagaluppi, O.~Darrigol, T.~Hartz, C.~Joas, A.~Kojevnikov, and O.~Pessoa (eds), \emph{The Oxford Handbook of the History of Quantum Interpretations} (Oxford: OUP), pp.~339--370. \url{https://arxiv.org/pdf/2011.09296} 

\

\noindent Leegwater, G. (2016), `An impossibility theorem for parameter independent hidden variable theories', \emph{Studies in History and Philosophy of Modern Physics} {\bf 54}, 18--34. \url{https://philsci-archive.pitt.edu/12067/1/CR_Phil-Sci.pdf}

\

\noindent Leifer, M. (2014), `Is the quantum state real? An extended review of $\psi$-ontology theorems', \emph
{Quanta} {\bf 3}(1), 67--155. \url{https://arxiv.org/pdf/1409.1570}

\

\noindent Maudlin, T. (1994), \emph{Quantum Non-Locality and Relativity: Metaphysical Intimations of Modern Physics} (Oxford: Blackwell; 3rd ed., Malden: Wiley-Blackwell, 2011).

\

\noindent Morgan, P. (2006), `Bell inequalities for random fields', \emph{ Journal of Physics A: Mathematical and General} {\bf 39}(23), 7441--7456. \url{https://arxiv.org/pdf/cond-mat/0403692}

\

\noindent Myrvold, W. (2016), `Lessons of Bell’s theorem: Nonlocality, yes; action at a distance, not necessarily', in M. Bell and S. Gao (eds), \emph{Quantum Nonlocality and Reality: 50 Years of Bell's Theorem} (Cambridge: CUP), pp.~238--260. \url{https://philsci-archive.pitt.edu/12382/1/MyrvoldBellCurrentVersionPhilSci.pdf}

\

\noindent Norsen, T. (2009), `Local causality and completeness: Bell vs. Jarrett', \emph{Foundations of Physics} {\bf 39}(3), 273--294. \url{https://arxiv.org/pdf/0808.2178}

\

\noindent Norsen, T., and Price, H. (2021), `Lapsing quickly into fatalism: Bell on backward causation', \emph{Entropy} {\bf 23}(2), 251/1--29. \url{https://www.mdpi.com/1099-4300/23/2/251} 

\

\noindent Palmer, T. (2020), `Discretization of the Bloch sphere, fractal invariant sets and Bell's theorem', \emph{Proceedings of the Royal Society A} {\bf 476}(2236), 20190350/1--24. \url{https://royalsocietypublishing.org/rspa/article/476/2236/20190350/80682}

\

\noindent de~la~Pe\~{n}a, L., and Cetto, A. M. (1996), \emph{The Quantum Dice: An Introduction to Stochastic Electrodynamics} (Dordrecht: Springer).

\

\noindent de~la~Pe\~{n}a, L., Cetto, A. M., and Vald\'{e}s-Hern\'{a}ndez, A. (2020), `Connecting two stochastic theories that lead to quantum mechanics', \emph{Frontiers in Physics} {\bf 8}, 162/1--8. \url{https://www.frontiersin.org/journals/physics/articles/10.3389/fphy.2020.00162/full}

\

\noindent Price, H. (1996), \emph{Time's Arrow \& Archimedes' Point: New Directions for the Physics of Time} (Oxford: OUP).

\

\noindent Schulman, L. (2012), `Experimental test of the ``special state'' theory of quantum measurement', \emph{Entropy} {\bf 14}, 665--686. \url{https://www.mdpi.com/1099-4300/14/4/665}  

\

\noindent Sen, I., and Valentini, A. (2020a), `Superdeterministic hidden-variables models I: Non-equilibrium and signalling', \emph{Proceedings of the Royal Society A} {\bf 476}(2243),  20200212/1--14. \url{https://royalsocietypublishing.org/rspa/article/476/2243/20200212/80953}

\

\noindent Sen, I., and Valentini, A. (2020b), `Superdeterministic hidden-variables models II: Conspiracy', \emph{Proceedings of the Royal Society A} {\bf 476}(2243), 20200214/1--14. \url{https://royalsocietypublishing.org/rspa/article/476/2243/20200214/80956}

\

\noindent Shimony, A. (1986), `Events and processes in the quantum world', in R. Penrose and C. Isham (eds), \emph{Quantum Concepts in Space and Time} (Oxford: OUP), pp.~182--203.

\

\noindent Shimony, A. (1989), `Search for a worldview which can accommodate our knowledge of microphysics', in J. Cushing and E. McMullin (eds),  \emph{Philosophical Consequences of Quantum Theory} (Notre Dame: University of Notre Dame Press), pp.~25--37.

\

\noindent Squires, E., Hardy, L., and Brown, H. (1994), `Non-locality from an analogue of the quantum Zeno effect', \emph{Studies in History and Philosophy of Science} {\bf 25}(3), 425--435.

\

\noindent Stuart, T., Slater, J., Colbeck, R., Renner, R., and Tittel, W. (2012), `An experimental test of all theories with predictive power beyond quantum theory', \emph{Physical Review Letters} {\bf 109}, 020402/1--5. \url{https://arxiv.org/pdf/1105.0133}

\

\noindent Valentini, A. (1991a), `Signal-locality, uncertainty, and the subquantum H-theorem. I', \emph{Physics Letters A} {\bf 156}(1--2), 5--11.

\

\noindent Valentini, A. (1991b), `Signal-locality, uncertainty, and the subquantum H-theorem. II', \emph{Physics Letters A} {\bf 158}(1--2), 1--8.

\

\noindent Valentini, A., (1992), `On the pilot-wave theory of classical, quantum and subquantum physics', PhD thesis, SISSA, Trieste. \url{https://iris.sissa.it/handle/20.500.11767/4334}

\

\noindent Valentini, A. (2002), `Signal-locality in hidden-variables theories', \emph{Physics Letters A} {\bf 297}(5--6), 273--278. \url{https://arxiv.org/pdf/quant-ph/0106098}

\

\noindent Valentini, A. (2024), `Pilot-wave theory and the search for new physics'. \url{https://arxiv.org/pdf/2411.10782}

\

\noindent Valentini, A., and Westman, H. (2005), `Dynamical origin of quantum probabilities', \emph{Proceedings of the Royal Society A} {\bf 461}(2053), 253--272. \url{https://arxiv.org/pdf/quant-ph/0403034}

\

\noindent Vervoort, L. (2013), `Bell's theorem: Two neglected solutions', \emph{Foundations of Physics} {\bf 43}(6), 769--791. \url{https://arxiv.org/pdf/1203.6587}

\

\noindent Vieira, C., Ramanathan, R., and Cabello, A. (2025), `Test of the physical significance of Bell non-locality', \emph{Nature Communications} {\bf 16}(1), 4390/1--7. \url{https://www.nature.com/articles/s41467-025-59247-7.pdf}

\

\noindent Wharton, K. (2014), `Quantum states as ordinary information', \emph{Information} {\bf 5}(1), 190--208. \url{https://www.mdpi.com/2078-2489/5/1/190}

\

\noindent Wharton, K., Sutherland, R., Amza, T., Liu, R., and Saslow, J. (2024), `A localized reality appears to underpin quantum circuits'. \url{https://arxiv.org/pdf/2412.05456}

\

\noindent Wilson, A. (2020), \emph{The Nature of Contingency: Quantum Physics as Modal Realism} (Oxford: OUP).

}

\end{document}